\documentclass{aa}  

\usepackage{upgreek}
\usepackage{txfonts}
\usepackage{booktabs} 
\usepackage{subcaption}
\usepackage{orcidlink}
\usepackage{hyperref}

\begin{document}

   \title{The GAPS programme at TNG LXIV:}
   \subtitle{An inner eccentric sub-Neptune and an outer sub-Neptune-mass candidate around BD+00\,444 (TOI-2443)}
   \titlerunning{An inner eccentric transiting sub-Neptune and an outer sub-Neptune-mass candidate around BD+00\,444.}

    \author{L. Naponiello\orcidlink{0000-0001-9390-0988}\inst{1}
          \and
          A.\,S. Bonomo\orcidlink{0000-0002-6177-198X}\inst{1} 
          \and
          L. Mancini\orcidlink{0000-0002-9428-8732}\inst{2,1,3} 
          \and
          M.-L. Steinmeyer\orcidlink{0000-0003-0605-0263}\inst{4} 
          \and
          K.~Biazzo\orcidlink{0000-0002-1892-2180}\inst{5} 
          \and
          D. Polychroni\orcidlink{0000-0002-7657-7418}\inst{1} 
          \and
          C. Dorn\orcidlink{0000-0001-6110-4610}\inst{4} 
          \and
          D. Turrini\orcidlink{0000-0002-1923-7740}\inst{1} 
          \and
          A.\,F. Lanza\orcidlink{0000-0001-5928-7251}\inst{6} 
          \and
          A. Sozzetti\orcidlink{0000-0002-7504-365X}\inst{1} 
          \and
          S. Desidera\orcidlink{0000-0001-8613-2589}\inst{7} 
          \and
          M. Damasso\orcidlink{0000-0001-9984-4278}\inst{1} 
          \and
          K.\,A. Collins\orcidlink{0000-0001-6588-9574}\inst{8} 
          \and
          I. Carleo\orcidlink{0000-0002-0810-3747}\inst{1} 
          \and
          K.\,I. Collins\orcidlink{0000-0003-2781-3207}\inst{9} 
          \and
          S. Colombo\orcidlink{0000-0002-3257-862X}\inst{10} 
          \and
          M.\,C. D'Arpa\orcidlink{0009-0004-5914-7274}\inst{11} 
          \and
          X. Dumusque\orcidlink{0000-0002-9332-2011}\inst{12} 
          \and
          M. Gonz\'alez\inst{13} 
          \and
          G. Guilluy\orcidlink{0000-0002-1259-2678}\inst{1} 
          \and
          V. Lorenzi\orcidlink{0000-0002-1958-9930}\inst{13,14} 
          \and
          G. Mantovan\orcidlink{0000-0002-6871-6131}\inst{15,7} 
          \and
          D. Nardiello\orcidlink{0000-0003-1149-3659}\inst{7} 
          \and
          M.~Pinamonti\orcidlink{0000-0002-4445-1845}\inst{1} 
          \and
          R.\,P. Schwarz\orcidlink{0000-0001-8227-1020}\inst{8} 
          \and
          V. Singh\orcidlink{0000-0002-7485-6309}\inst{5} 
          \and
          C.\,N. Watkins\orcidlink{0000-0001-8621-6731}\inst{8} 
          \and
          T. Zingales\orcidlink{0000-0001-6880-5356}\inst{15,7} 
          }

    \institute{INAF -- Osservatorio Astrofisico di Torino,
              Via Osservatorio 20, 10025 Pino Torinese, Italy\\
              \email{luca.naponiello@inaf.it}
            \and 
            Dipartimento di Fisica, Università degli studi di Roma ``Tor Vergata'', Rome, Italy
            \and 
            Max-Planck-Institut für Astronomie, Heidelberg, Germany
            \and 
            Institute for Particle Physics and Astrophysics, ETH Zürich, Otto-Stern-Weg 5, 8093 Zürich, Switzerland
            \and 
            INAF -- Osservatorio Astronomico di Roma, Monte Porzio Catone, Italy
            \and 
            INAF -- Osservatorio Astrofisico di Catania, Via S. Sofia 78, 95123 Catania, Italy
            \and 
            INAF -- Osservatorio Astronomico di Padova, Padova, Italy
            \and 
            Center for Astrophysics \textbar \ Harvard \& Smithsonian, 60 Garden Street, Cambridge, MA 02138, USA
            \and 
            George Mason University, 4400 University Drive, Fairfax, VA, 22030 USA
            \and 
            INAF -- Osservatorio Astronomico di Palermo, Piazza del Parlamento 1, 90134, Palermo, Italy
            \and 
            Dipartimento di Fisica e Chimica ``Emilio Segrè'', Università di Palermo, Via Archirafi 36, Palermo, Italy
            \and 
            Observatoire de Gen\'{e}ve, Universit\'{e} de Gen\'{e}ve, 1290 Versoix, Switzerland
            \and 
            INAF -- Fundación Galileo Galilei, Rambla J. A. F. Pérez 7, 38712 Breña Baja, TF, Spain
            \and 
            Instituto de Astrofísica de Canarias (IAC), c/ Vía Láctea s/n, 38205, La Laguna (Tenerife), Canary Islands, Spain
            \and 
            Dipartimento di Fisica e Astronomia ``Galileo Galilei'', Vicolo dell’Osservatorio 3, 35122 Padova, Italy
             }

   \date{Submitted 12 August 2024; Accepted 13 November 2024.}

 
  \abstract
   {Super-Earths and sub-Neptunes are the most common types of planets outside the Solar System and likely represent the link between terrestrial planets and gas giants. Characterizing their physical and orbital properties and studying their multiplicity are key to testing and understanding their formation, migration, and evolution.}
   {We examined in depth the star BD+00\,444 (GJ\,105.5, TOI-2443; $V=9.5$\,mag; $d=23.9$\,pc), with the aim of characterizing and confirming the planetary nature of its small companion, the planet~candidate TOI-2443.01, which was discovered by the \emph{TESS} space telescope and subsequently validated by a follow-up statistical study.}
   {We monitored BD+00\,444 with the HARPS-N spectrograph for 1.5 years to search for planet-induced radial-velocity (RV) variations, and then analyzed the RV measurements jointly with \emph{TESS} and ground-based photometry.}
   {We determined that the host is a quiet K5\,V star with a radius of $R_{\star}=0.631^{+0.013}_{-0.014}\,R_{\sun}$ and a mass of $M_{\star}=0.642^{+0.026}_{-0.025}\,M_{\sun}$. We revealed that the sub-Neptune BD+00\,444\,b has a radius of $R_{\rm b}=2.36\pm0.05\,R_{\oplus}$, a mass of $M_{\rm b}=4.8\pm1.1\,M_{\oplus}$ and, consequently, a rather low-density value of $\rho_{\rm b}=2.00^{+0.49}_{-0.45}$ g\,cm$^{-3}$, which makes it compatible with both an Earth-like rocky interior with a thin H-He atmosphere and a half-rocky, half-water composition with a small amount of H-He. Having an orbital period of about 15.67 days and an equilibrium temperature of about 519 K, BD+00\,444\,b has an estimated transmission spectroscopy metric (TSM) of $159^{+46}_{-31}$, which makes it ideal for atmospheric follow-up with the James Webb Space Telescope. Notably, it is the second most eccentric inner transiting planet, $e=0.302^{+0.051}_{-0.035}$, with a mass below $20\,M_{\oplus}$, among those with well-determined eccentricities. We estimated that tidal forces from the host star affect both planet~b’s rotation and eccentricity, and strong tidal dissipation may signal intense volcanic activity.
   Furthermore, our analysis suggests the presence of a sub-Neptune-mass planet~candidate, BD+00\,444\,c, having an orbital period of $P_{\rm c}=96.6\pm1.4$\,days, and a minimum mass $M_{\rm c}\sin{i}=9.3^{+1.8}_{-2.0}\,M_{\oplus}$. With an equilibrium temperature of about 283 K, BD+00\,444\,c is right inside the habitable zone; however, this candidate necessitates further observations and stronger statistical evidence to be confirmed. We explored the formation and migration of both planets by means of population synthesis models, which revealed that both planets started their formation beyond the water snowline during the earliest phases of the life of their protoplanetary disk.
  }
   {}


   \maketitle
%

\section{Introduction}
One of the main findings of the NASA Kepler mission \citep{Borucki2010} is that, in our Galaxy, small planets ($R_{\rm p}<4\,R_{\oplus}$) at short orbital periods ($P_{\rm orb}<100$\,days) are the most common type (see, e.g., \citealt{Bean2021, Biazzo2022a} for recent and comprehensive reviews). Small planets with $1R_{\oplus}<R_{\rm p}<4\,R_{\oplus}$ are thought to bridge the gap between the rocky, terrestrial planets and gas giants, even though they may have very large masses, such as the ultra-dense K2-292\,b \citep[$M\approx25\,M_{\oplus}$;][]{Luque2019}. 
{\it Kepler} also revealed a deficit in the occurrence rate distribution at 1.5--2.0\,$R_\oplus$, which is associated with a bimodality in their radius distribution \citep{Fulton2017}, probably depending on whether their rocky cores have been able to retain a substantial hydrogen-rich envelope or not. This dichotomy has now become the classical way to a further sub-classification into super-Earths and sub-Neptunes. At the same time, statistical studies of the orbital eccentricity of transiting planets with $R_{\rm p}<6\,R_{\oplus}$ tell us that single transiting planets display higher average eccentricities than transiting planets in multiple systems, possibly reflecting differences in their formation pathways \citep{VanEylen2015,Xie2016}.
Numerical simulations show that massive external companions in multi-planet systems containing sub-Neptunes may perturb their orbits by exciting their eccentricity, producing wider spacing and increasing the mutual inclination among planets \citep{Pu2018}, thus making it more unlikely to observe multiple transiting planets. This result can explain why systems with fewer transiting planets are dynamically hotter (i.e., they are more eccentric and inclined) than those with more transiting planets, although this excitation can also be self-induced in tightly packed systems \citep{VanEylen2019}.

Currently, with \emph{TESS} \citep{Ricker2014} continuously scanning the skies, a large number of new sub-Neptune exoplanets are being discovered (see, e.g., \citealt{2020AJ....160..113B,2021A&A...650A.145C,2021AJ....162...87B,2022MNRAS.514.4120G,2023A&A...670A.136K,2024A&A...685A..56S,2024A&A...687A.226M}). Characterizing these planets is crucial for enhancing our understanding of this class of exoplanets, particularly those that are favorable targets for atmospheric studies with the JWST. Indeed, observational constraints on sub-Neptune atmospheres can provide information on their compositions, though their atmospheric chemistry is still largely unknown given the technical difficulty of probing them until recently \citep[e.g.,][]{Kreidberg2014,Tsiaras2016,Mikal-Evans2023}. Thanks to JWST, which began operations in July 2022, we just entered a new era of atmospheric characterization \citep{JWST2023}. The first investigations of sub-Neptunes atmospheres have already been performed, like those of K2-18\,b \citep{Madhusudhan2023} and TOI-270\,d \citep{Holmberg2024}, although the interpretation of these transmission spectra raised some controversies (e.g., \citealt{Wogan2024}).
A large observational program is currently running to investigate the atmospheric presence and composition of a dozen super-Earths and sub-Neptunes exoplanets \citep[i.e., Compositions of Mini-Planet Atmospheres for Statistical Study, COMPASS;][]{Wallack2024}.

\begin{figure}
\centering
\includegraphics[width=0.45\textwidth]{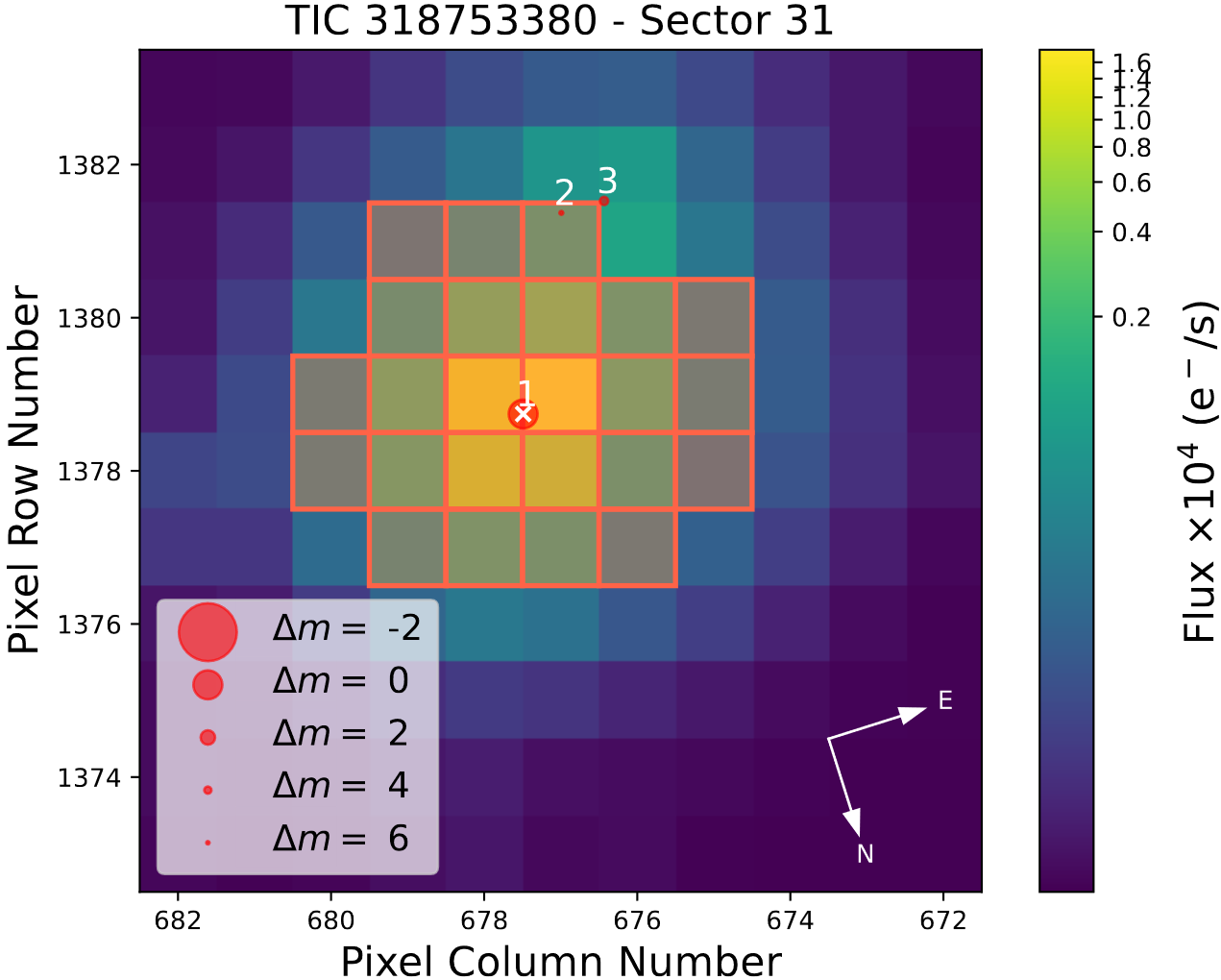}
\caption{Target pixel file from the \emph{TESS} observation of Sector~31, centered on BD+00\,444, which is marked with a white cross. The SPOC pipeline aperture is shown by shaded red squares, and the {\it Gaia} satellite DR3 catalog \citep{Gaia2023} is also overlaid with symbol sizes proportional to the {\it Gaia} magnitude difference with BD+00\,444.}
\label{fig:star}
\end{figure}
On 6 January 2021, the \emph{TESS} target star TIC\,318753380 (BD+00\,444, GJ\,105.5) was officially named TOI-2443 (\emph{TESS} Object of Interest; \citealt{Guerrero2021}), following the summary of the data validation report (DVR) produced by the \emph{TESS} Science Processing Operations Center (SPOC) \citep{Jenkins2016} pipeline at the NASA Ames Research Center through the Transiting Planet Search (TPS; \citealt{jenkins2002,jenkins2010,Jenkins2020b}) and data validation (\citealt{Twicken2018}, \citealt{Li2019}) modules. In particular, as stated in this report, the difference image centroid locates the source of the transits within $4$ arcsec of the target star’s location. The candidate TOI-2443.01 was designated as planet TOI-2443\,b after the statistical validation of \citet{Mistry2023}, in which the false positive probability was evaluated using the {\tt TRICERATOPS} tool \citep{Giacalone2021}. However, in this work, we use the star's original designation and refer to the planet as BD+00\,444\,b.

\begin{figure*}
\centering
\includegraphics[width=0.9\textwidth]{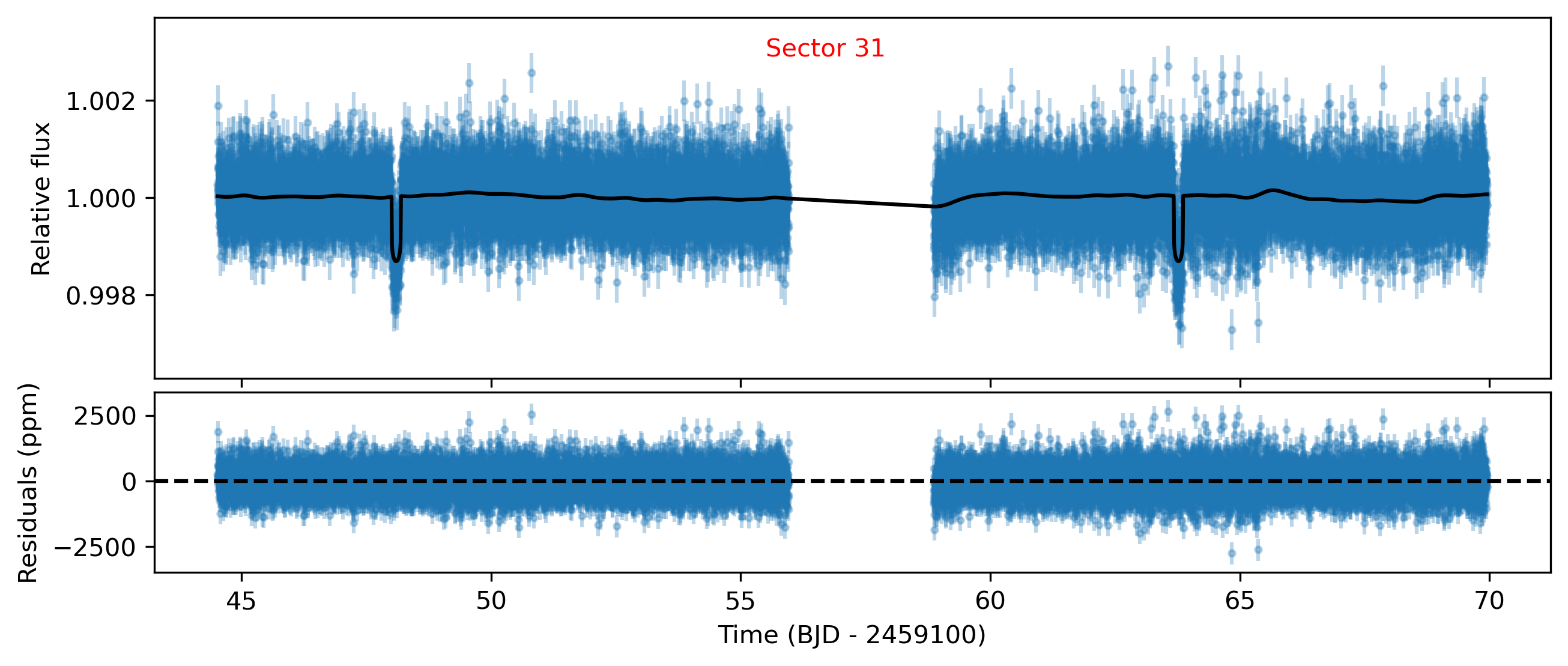}
\caption{Light curve of BD+00\,444 as collected by \emph{TESS} in Sector~31 with a two-minute cadence. {\it Top panel:} Light curve from the PDC-SAP pipeline. The black line represents our best-fit transit model (from Sect.\,\ref{sec:joint_analysis}). {\it Bottom panel:} Residuals of the best-fit model in parts per million.}
\label{fig:lc_trend}
\end{figure*}

Within the Global Architecture of Planetary Systems (GAPS) Neptune project \citep{Naponiello2022,Naponiello2023,Damasso2023}, we started monitoring this target in July 2021, as the potential host of a sub-Neptune exoplanet. In particular, we took Doppler measurements with the High Accuracy Radial velocity Planet Searcher for the Northern hemisphere (HARPS-N; \citealt{Cosentino2012}) instrument installed at the Telescopio Nazionale Galileo (TNG) in La Palma (Spain). Here, we report the results of our observations and analyses that allowed us to measure the mass of BD+00\,444\,b and discover a new candidate planet in the system. In particular, BD+00\,444\,b emerged as the second most eccentric inner transiting planet with a mass below $20M_{\oplus}$, among those with eccentricities determined with at least 5$\sigma$ accuracy, after the recently discovered TOI-757\,b \citep{Alqasim2024}, and the only one with a possible companion in the habitable zone (Sect.\,\ref{sec:candidate}). 

This paper is organized as follows. Section\,\ref{sec:data_reduction} contains the details of the photometric and RV observations. The star characterization is detailed in Sect.\,\ref{sec:star_char}, while the analysis of the planetary signals is presented in Sect.\,\ref{sec:analysis} and discussed in Sect.\,\ref{sec:discussion}. Finally, we draw the conclusions in Sect.\,\ref{sec:conclusions}.

\section{Observations and data reduction}
\label{sec:data_reduction}
\subsection{TESS photometry}
\label{sec:photometry}

The pre-selected target BD+00\,444, located at about 23.9\,pc, was shortly observed by \emph{TESS} in Sector~31, between 21 October and 19 November 2020, and only two transit events were recorded. The two-minute cadence photometry of BD+00\,444 from \emph{TESS} spans about 25 days, with a gap of about 4 days in-between, and it was analyzed with the python wrapper \texttt{juliet}\footnote{\texttt{\url{https://juliet.readthedocs.io/en/latest/}}} \citep{Espinoza2019} (as detailed in Sect.\,\ref{sec:joint_analysis}). In particular, we used the Presearch Data Conditioning Simple Aperture Photometry (PDC-SAP; \citealt{Stumpe2012, Stumpe2014}, \citealt{Smith2012}) light curve, which is provided by \emph{TESS} SPOC pipeline and retrieved via the Python package \texttt{lightkurve} \citep{lightkurve} from the Mikulski Archive for Space Telescopes. The PDC-SAP photometry is already corrected for dilution from other objects contained within the aperture using the Create Optimal Apertures module \citep{Bryson2010, Bryson2020}. In this case there are only two sources close to the borders of the SPOC aperture, within a 5--6\,mag difference, as shown in Fig.~\ref{fig:star} using \texttt{tpfplotter}\footnote{\texttt{tpfplotter} is a python package developed by J. Lillo-Box and publicly available on \url{www.github.com/jlillo/tpfplotter}.} \citep{Aller2020}. The PDC-SAP light curve is plotted, along with the two transits, in Fig.~\ref{fig:lc_trend}.

\subsection{LCOGT light curve follow up}
Two near-full transits of BD+00\,444\,b were observed from two sites of the Las Cumbres Observatory Global Telescope \citep[LCOGT;][]{Brown2013} 1\,m network. The first transit was observed on 19 September 2021 in the Pan-STARRS $Y$ band ($\lambda_{\rm c} = 10040$\,\AA, ${\rm \Delta \lambda} =1120$\,\AA) from the Siding Spring Observatory near Coonabarabran, Australia. The second transit was observed on 10 September 2023 in Pan-STARRS $z_s$ band ($\lambda_{\rm c} = 8700$\,\AA, ${\rm \Delta \lambda} =1040$\,\AA) from the McDonald Observatory near Fort Davis, Texas, USA. The LCOGT telescopes are equipped with a $4096\times4096$ SINISTRO camera having an image scale of $0\farcs389$ per pixel, resulting in a $26\arcmin\times26\arcmin$ field of view, and the images were calibrated using the standard LCOGT {\tt BANZAI} pipeline \citep{McCully2018}, while differential photometric data were extracted using {\tt AstroImageJ} \citep{Collins2017}. We used circular photometric apertures with radii $5\farcs0$--$7\farcs8$ that excluded all the flux from the nearest known neighbors in the {\it Gaia} DR3 catalog (Fig.~\ref{fig:star}). The light curve data are available on the {\tt EXOFOP-TESS} website\footnote{\url{https://exofop.ipac.caltech.edu/tess/target.php?id=318753380}} and are included in the global modeling described in Sect. \ref{sec:joint_analysis}.

\subsection{HARPS-N high-resolution spectra}
\label{sec:rv}

Between 18 July 2021 and 18 January 2023, we collected a total of 97 high-resolution spectra of BD+00\,444 with HARPS-N (Table\,\ref{tab:rv}) with a variable exposure time of 15--20\,min. The RVs and activity indices were extracted from the spectra reduced using version 3.0.1 of the HARPS-N Data Reduction Software ({\tt DRS}, \citealt{Dumusque2021}), available on the Data Analysis Center for Exoplanets (DACE) web platform. In the following analysis, we employed all the RVs, including a few with relatively lower signal-to-noise ratio (S/N), as their removal did not influence the results. Overall, the measurements have an average error of $1.6$\,m\,s$^{-1}$, a root mean square error of $3.3$\,m\,s$^{-1}$, and an S/N\,$\approx$\,70, measured at a reference wavelength of 5500 {\AA}.

\section{Host star characterization}
\label{sec:star_char}
\subsection{Stellar parameters}
\label{sec:star}
A co-added spectrum was produced to derive stellar atmospheric parameters (effective temperature $T_{\rm eff}$, surface gravity $\log g_{\star}$, micro-turbulence velocity $\xi$, and iron abundance [Fe/H]) through the \texttt{MOOG} code (\citealt{sneden1973}, version 2019) and adopting the MARCS atmospheric models \citep{Gustafsson2008}. Similar results were obtained using the ATLAS9 grid of model atmospheres with new opacities (\citealt{castellikurucz2003}).
We adopted the list of \ion{Fe}{i} and \ion{Fe}{ii} lines by \cite{Biazzo2022} and imposed the condition that the \ion{Fe}{i} abundance does not depend on the excitation potential and equivalent width of the lines for deriving $T_{\rm eff}$ and $\xi$, respectively, and the \ion{Fe}{i}/\ion{Fe}{ii} ionization equilibrium for determining $\log g$. Our spectroscopic analysis provides as the final atmospheric parameters those listed in Table\,\ref{tab:star}, in particular: $T_{\rm eff}=4375\pm65$ K, $\log g_{\star}=4.42\pm0.20$ cgs, $\xi=0.35\pm0.35$ km\,s$^{-1}$, and [Fe/H]$=-0.29\pm0.12$ dex. 

We also estimated the effective temperature considering {\it Gaia} DR3 \citep{Gaia2016, Gaia2023} and 2MASS photometry \citep{Skrutskie2006} and using the \texttt{colte} tool \citep{Casagrande2021} with reddening taken from \cite{Capitanio2017}. The total extinction coefficient, $A_V$, is consistent with zero, as expected from a star at only about 24\,pc. The final weighted average of the photometric $T_{\rm eff}$ from twelve colors is $4376\pm81$\,K, in excellent agreement with our spectroscopic result. The stellar projected rotational velocity ($v\sin{i} $) was obtained through the 2009 version of the {\tt MOOG} code \citep{sneden1973} and the spectral-synthesis technique of three regions around 5400, 6200, and 6700\,{\AA} \citep{Biazzo2022}. By assuming a macroturbolence velocity \citep{Brewer2016} $v_{\rm macro}=1.4$\,km\,s$^{-1}$, we found $v\sin{ i}=1.7\pm0.8$\,km\,s$^{-1}$, which is basically at the HARPS-N resolution limit, thus suggesting an inactive, slowly rotating star (see also Sect.\,\ref{sec:age}), unless it is observed nearly pole-on.

We determined the stellar physical parameters, i.e. mass, radius, and age, with a Bayesian differential evolution Markov chain Monte Carlo framework through the {\tt EXOFASTv2} tool \citep{2017ascl.soft10003E}. 
Specifically, we performed simultaneous modeling of the stellar spectral energy distribution (SED)  and the MESA Isochrones and Stellar Tracks (MIST, \citealt{2015ApJS..220...15P}) by maximizing a combined Gaussian likelihood function given by the product of the SED and the MIST likelihoods (see \citealt{Eastman2019} for more details). We sampled the SED with the Tycho-2 $B_T$ and $V_T$, 2MASS $J$, $H$, and $K_s$, and WISE $W1$, $W2$, $W3$, and $W4$ magnitudes (see Table~\ref{tab:star})\footnote{We did not use the {\it Gaia} magnitudes because of their wide bandpasses, following the suggestion of \citet{Eastman2019}.}; we modeled it with the NexGen stellar atmospheric models \citep{Allard2012} by varying $T_{\rm eff}$, $\log g_{\star}$, [Fe/H], the ratio of the stellar radius to the distance ($R_\star / d$), and the $V$-band extinction $A_V$. Two additional free parameters for the MIST stellar models, properly interpolated by {\tt EXOFASTv2}, are the stellar mass ($M_\star$) and age. 
We made use of previous information by imposing  Gaussian priors on the  $T_{\rm eff}$ and [Fe/H] as derived from the spectral analysis as well as on the {\it Gaia} DR3 parallax.
The fitted and derived (e.g., stellar luminosity and density) parameters of the host star are given in Table~\ref{tab:star}. Our stellar parameters supersede a number of previous determinations in the literature (e.g., \citealt{Mishenina2008, Yee2017}).


\begin{table}
\centering %
\caption{Stellar parameters of BD+00\,444.} %
\label{tab:star} %
\resizebox{\hsize}{!}{
\begin{tabular}{l c c  c}
\hline %
\hline  \\[-8pt]
Parameter & Unit & Value & Source \\
\hline  \\[-6pt]
\multicolumn{1}{l}{\large{Identifiers}} \\ [2pt] %
TOI \dotfill     & \dotfill& 2443 & TOI catalog \\
TIC \dotfill & \dotfill&  318753380 & TIC \\
HIP \dotfill & \dotfill&  12493 & HIP \\
GJ \dotfill & \dotfill&  105.5 & GJ \\
Tycho-2 \dotfill   & \dotfill&  0047-00611-1 & Tycho-2 \\
2MASS \dotfill & \dotfill& J02404288+0111554 & 2MASS \\
{\it Gaia}  \dotfill & \dotfill& 2501948402746099456 & {\it Gaia}~DR3 \\ [6pt] %
\multicolumn{1}{l}{\large{Astrometric properties}} \\ [2pt] %
$\alpha$\,(J2016.0) \dotfill & h & 02:40:43.18  & {\it Gaia}~DR3 \\
$\delta$\,(J2016.0)  \dotfill & deg & +01:11:58.95  & {\it Gaia}~DR3 \\
$\pi$ \dotfill & mas & $41.822 \pm 0.021$ & {\it Gaia}~DR3 \\
$\mu_\alpha \cos{\delta}$ \dotfill & mas\,yr$^{-1}$  & $283.915\pm0.024$ & {\it Gaia}~DR3 \\
$\mu_\delta$ \dotfill & mas\,yr$^{-1}$  & $231.74\pm0.02$ & {\it Gaia}~DR3 \\
$d$ \dotfill & pc  & $23.911\pm0.011$ & This work \\ [6pt] %
\multicolumn{1}{l}{\large{Photometric properties}} \\ [2pt] %
$B_{\rm T}$ \dotfill & mag & $11.024\pm0.045$ & Tycho-2 \\ 
$V_{\rm T}$ \dotfill & mag & $9.674\pm 0.023$ & Tycho-2 \\ 
$BP$ \dotfill & mag & $9.7283\pm0.0011$ & {\it Gaia}~DR3 \\
$G$ \dotfill & mag & $9.0378\pm0.0003$ & {\it Gaia}~DR3 \\
$RP$ \dotfill & mag & $8.2339\pm0.0006$ & {\it Gaia}~DR3 \\
$J$ \dotfill & mag & $7.260\pm0.021$ & 2MASS \\
$H$ \dotfill & mag & $6.692\pm0.046$ & 2MASS \\
$K_{\rm S}$ \dotfill  & mag & $6.500\pm0.018$ & 2MASS \\ 
$W1$ \dotfill & mag & $6.443\pm0.072$ & AllWISE \\
$W2$ \dotfill & mag & $6.422\pm0.026$ & AllWISE \\     
$W3$ \dotfill & mag & $6.453\pm0.016$ & AllWISE \\ 
$W4$ \dotfill & mag & $6.378\pm0.051$ & AllWISE \\ 
$A_V$ \dotfill & mag & $<0.091$ & This work \\ [6pt] %
\multicolumn{1}{l}{\large{Stellar parameters}} \\ [2pt] %
Spectral type \dotfill & & K5\,V & \citet{Stephenson1986} \\
$L_{\star}$ \dotfill & $L_{\sun}$ & $0.1365\pm0.0056$ & This work \\
$M_{\star}$ \dotfill & $M_{\sun}$ & $0.642\pm0.026$ & This work \\
$R_{\star}$ \dotfill & $R_{\sun}$ & $0.631\pm0.014$ & This work \\
$T_{\rm eff}$ \dotfill & K & $4375\pm65$ & This work\\ 
$v\sin{i}$ \dotfill & km\,s$^{-1}$ & 1.7 $\pm$ 0.8 & This work \\
$\log g_{\star}$ \dotfill & cgs & $4.42\pm0.20$ & This work \\
$\xi$ \dotfill & km\,s$^{-1}$ & $0.35\pm0.35$ & This work \\
$\rm{[Fe/H]}$ \dotfill & dex & $-0.29\pm0.12$ & This work \\
$\rho_{\star}$\dotfill & g\,cm$^{-3}$ & $3.61\pm0.24$ & This work \\
$\log R^{\prime}_{\rm HK}$\dotfill & dex &  $-5.004\pm0.006$ & This work \\
Age\dotfill & Gyr & $7.5^{+4.3}_{-4.6}$ & This work \\
$U^{(a)}$ \dotfill & km\,s$^{-1}$ & $77.6$ & This work \\ 
$V^{(a)}$ \dotfill & km\,s$^{-1}$ & $6.1$ & This work \\ 
$W^{(a)}$ \dotfill & km\,s$^{-1}$ & $-31.4$ & This work \\ 
\hline %
\end{tabular}
}
\tablebib{TESS Primary Mission TOI catalog \citep{Guerrero2021}; TIC \citep{Stassun2018,Stassun2019}; HIP \citep{Perryman1997}; GJ \citep{Gliese1991}; Tycho-2 \citep{hog}; 2MASS \citep{Skrutskie2006}; {\it Gaia} DR3 \citep{Gaia2023}; AllWISE \citep{Allwise2013}.}
\tablefoot{
\tablefoottext{a}{Galactic velocity components, where $U$ is positive towards the Galactic anti-centre, $V$ in the direction of the Galactic rotation, and $W$  towards the Galactic North Pole}.}
\end{table}

\subsection{Stellar age indicators}
\label{sec:age}

We can estimate the age of our target using its rotation period (gyro-chronology), its chromospheric activity, or its spatial velocity components. The modulation of chromospheric lines might suggest a rotation period of about 45\,d (Sect.~\ref{sec:periodogram}) that can be used with the rotational isochrones of \citet{Barnesetal16} assuming a Johnson $B-V$ color of $1.20$\,mag \citep{Koen2010}. The precision in stellar age is limited by the dispersion in the relationship between rotation, age, and color index. Additionally, the subsolar metallicity of our target is expected to reduce magnetic braking efficiency \citep{Amardetal20}. Therefore, its true age can be older than estimated from the above standard gyro-chronological relationship. 

The chromospheric activity index of BD+00\,444, $\log R^{\prime}_{\rm HK}$, is equal to $-5.004 \pm 0.006$, and it can be used to estimate old stellar ages according to \citet{MamajekHillebrand08}. Given the sub-solar metallicity of our star, its chromospheric index is also expected to be enhanced at a given age with respect to a star with solar metallicity \citep{Rocha-PintoMaciel98}. The $\log R^{\prime}_{\rm HK}$ was extracted from the DACE web interface\footnote{\url{https://dace.unige.ch/dashboard/}}, and is computed from the $S$-index following \citet{Noyes1984}. The $S$-index was derived by extracting the emission in the center of the Ca~{\sc ii} H and K lines from order-merged spectra using a triangular band-pass normalized by two pseudo-continua on the blue and red-side of the two spectral lines, as defined by \citet{Vaughan1978}. The corresponding $\log R^{\prime}_{\rm HK}$ derived is therefore compatible with the original Mount-Wilson survey measurements \citep{Wilson1978}.

We computed the Galactic velocity components of BD+00\,444 with respect to the Sun (Table\,\ref{tab:star}) using the method of \citet{JohnsonSoderblom87} as implemented in the IDL Astrolib procedure {\tt gal\_uvw.pro}. Adopting the $UVW$ method of \citet{Almeida-FernandesRocha-Pinto18}, we estimated a kinematical age of $7.8$~Gyr with a typical uncertainty of at least $3.0$~Gyr, consistent with the age estimation from the stellar evolutionary tracks (cf.~Table\,\ref{tab:star}). Nevertheless, we recall that kinematically based ages are highly uncertain for single stars because it is possible to find stars with an age of 1--3\,Gyr having $U, V, W$ velocity components typical of stars with ages of the order of 7--10\,Gyr. In other words, the kinematical age is merely suggestive for a single star and provides a statistically meaningful indication only for a sufficiently large sample of coeval stars without kinematic peculiarities.

In conclusion, all the indirect methods considered above indicate that BD+00\,444 is likely to be older than the Sun, although with a large uncertainty that makes its true age fall between about 1--2 and 10~Gyr.

\subsection{Stellar companions}
\label{sec:binarity}

The high-contrast follow-up detailed by \citet{Mistry2023}, including speckle imaging techniques with high-resolution observations, did not reveal the presence of any stellar companion. Similarly, despite possible offsets between the different instruments, we did not find significant long-term RV variations between the average RV from the HARPS-N orbital solution (Sect.\,\ref{sec:analysis}) and previous measurements: $72986.35^{+0.24}_{-0.19}$ m\,s$^{-1}$ from HARPS-N ($N$=97 measurements between 2021 and 2023), $72930\pm130$ m\,s$^{-1}$ from {\it Gaia} DR3 (2014--2017), $72926\pm3$ m\,s$^{-1}$ from SOPHIE \citep{Perruchot2008} ($N$=1, 2011), $72760\pm50$ m\,s$^{-1}$ from ELODIE \citep{Baranne1996}($N$=3, 2000-2005). The {\it Gaia} DR3 archive also shows no co-moving objects within $900''$ or $\sim$20\,000\,au (i.e. no star with comparable parallax or proper motion). In the {\it Hipparcos}-{\it Gaia} astrometric acceleration (also known as proper motion anomaly, PMA) catalogs \citep{Brandt2021,Kervella2022}, BD+00\,444 was reported with no statistically significant indication of binarity, placing rather stringent constraints on the presence of low-mass giant planets ($\approx50$ $M_\oplus$) in the 3--10 au separation regime. However, BD+00\,444 in the {\it Gaia} DR3 archive is reported to have a Renormalized Unit Weight Error (RUWE) of 1.25. This diagnostic is typically used to identify departures from a satisfactory single-star fit that might indicate the presence of orbital motion in {\it Gaia} DR3 astrometry (e.g., \citealt{Lindegren2021}), with a RUWE threshold of 1.4 usually adopted to discriminate the two categories. For nearby ($\varpi>35$ mas), bright ($G<12$ mag) stars with {\it Gaia} color in the range $1$\,mag\,$<BP-RP<2$\,mag as is the case for BD+00\,444, the typical RUWE is close to 1.0 \citep{Sozzetti2023}. In principle, the presence of a giant planetary companion orbiting BD+00\,444 at 1--3 au might help explain the (mild) departure from RUWE $=1.0$ for the star, as in this range of orbital separations the PMA technique begins to lose sensitivity due to the smearing effect of the orbital motion for periods comparable to or shorter than the duration of the {\it Hipparcos} and {\it Gaia} DR3 observations. However, as shown later on in Section \ref{sec:detection}, a gas giant at such orbital distances should have been readily spotted based on the HARPS-N RV data at hand, unless it were to lie on a low inclination orbit (a detailed investigation of such a scenario from a dynamical stability standpoint is beyond the scope of this paper). A possible alternative explanation for the moderately elevated RUWE is instead due to larger unmodeled systematics in the astrometric time series. 

\begin{figure*}
  \centering
  \begin{subfigure}{0.33\linewidth} 
    \centering
    \includegraphics[width=\linewidth]{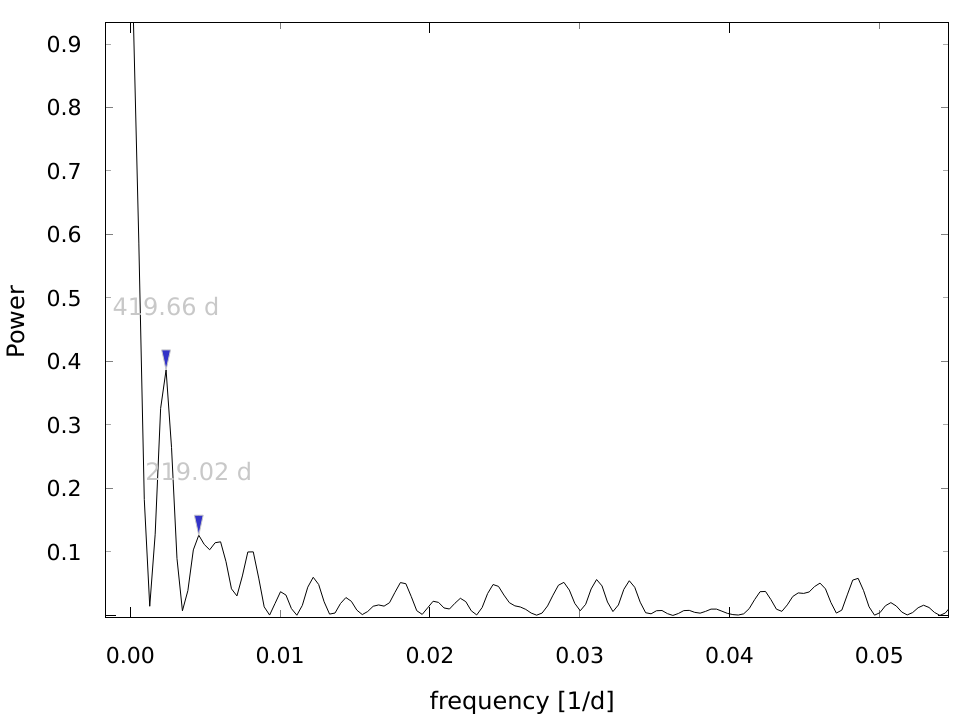}
  \end{subfigure}
  \hfill
  \begin{subfigure}{0.33\linewidth}
    \centering
    \includegraphics[width=\linewidth]{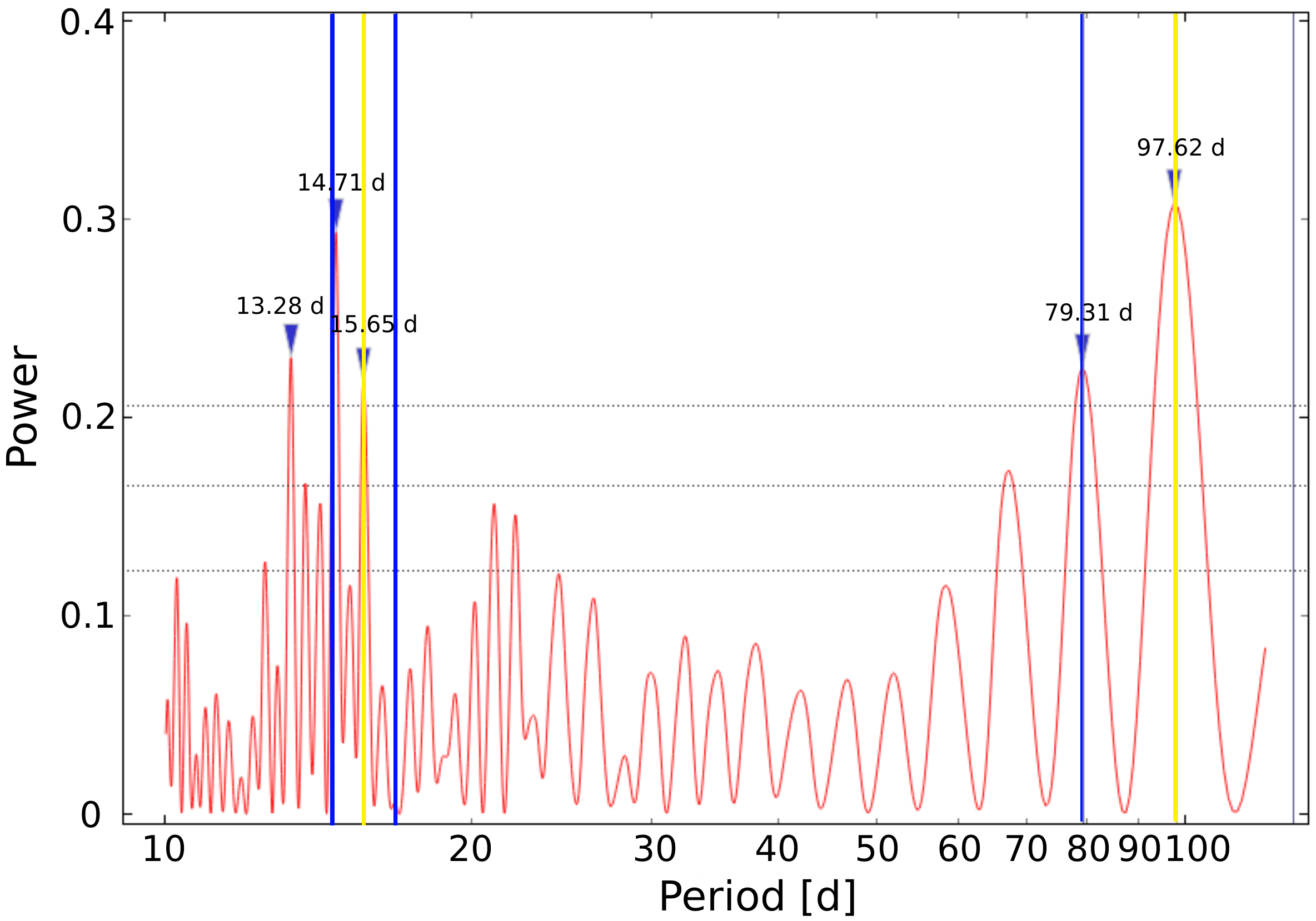}
  \end{subfigure}
  \hfill
  \begin{subfigure}{0.33\linewidth}
    \centering
    \includegraphics[width=\linewidth]{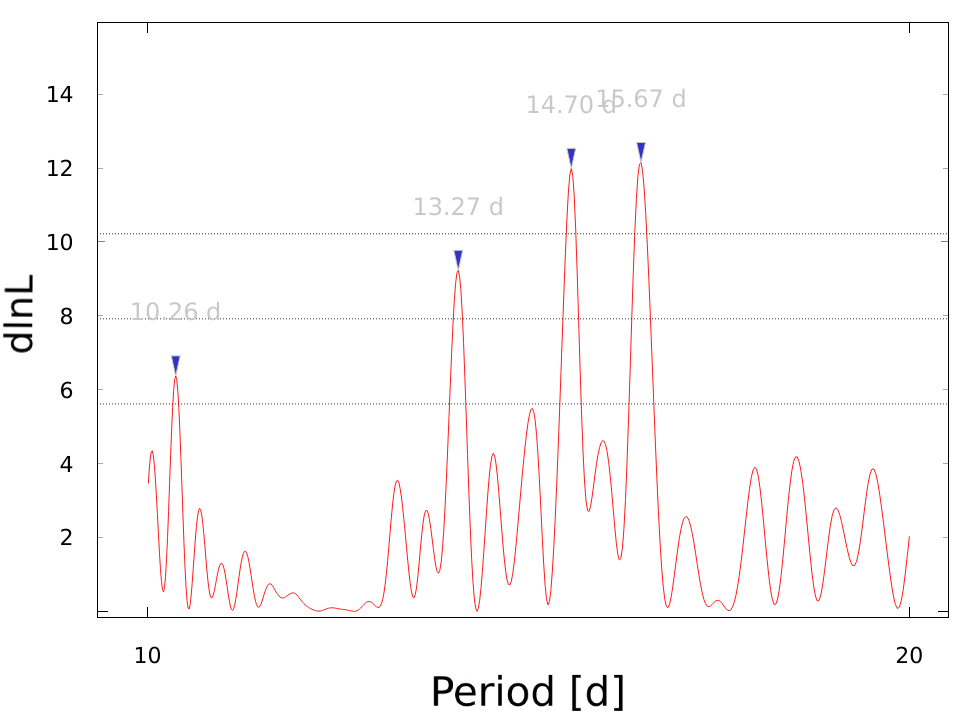}
  \end{subfigure}
  \vspace{1em} 
  \begin{subfigure}{0.85\textwidth}
    \centering
    \includegraphics[width=\linewidth]{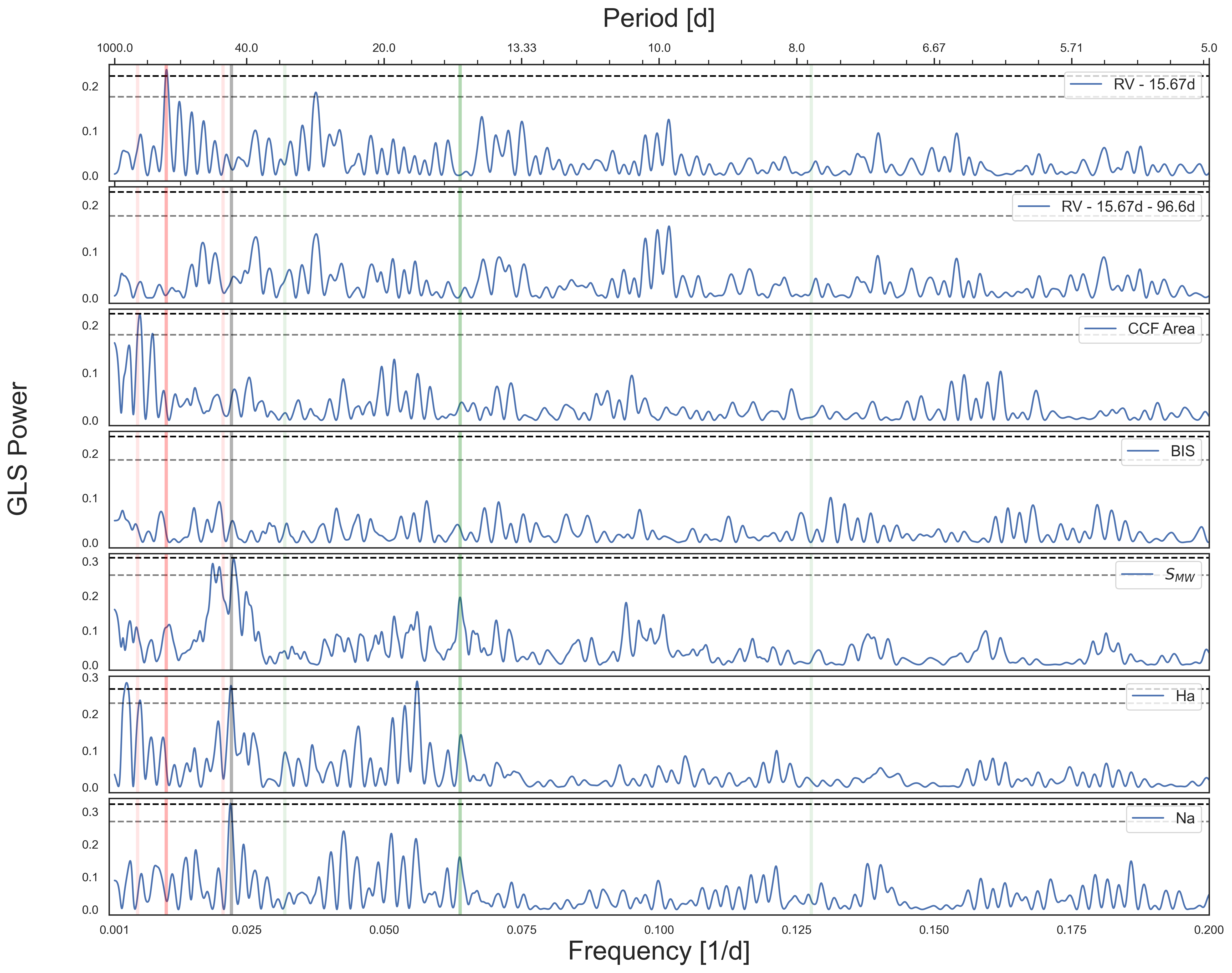}
  \end{subfigure}
  \caption{
    \textbf{Top row, left panel}: The window function of HARPS-N RVs, with its highest peaks highlighted. 
    \textbf{Top row, middle panel}: The GLS of HARPS-N RVs. Two yellow vertical lines are drawn at the periodicity of BD+00\,444\,b transits (15.67\,days) and the highest peak of the GLS, while their aliases due to the window function (respectively $\approx \,219$ and $420$\,days), are identified by blue vertical lines. 
    \textbf{Top row, right panel}: The MLP of HARPS-N RVs between 10 and 20 days, with its major peak identified at precisely 15.67 days. The three horizontal lines represent the FAP levels of 10\%, 1\%, and 0.1\%. 
    \textbf{Bottom panel}: GLS periodogram of the HARPS-N RV residuals of the 1-planet and 2-planet models (first two rows), and of various activity indices specified in the labels. The main peak of the RV GLS periodogram, which is here identified as the signal of planet~candidate BD+00\,444\,c at about $97$~d, is highlighted with a red vertical line, while the period of the transiting candidate (15.67\,d) and the possible stellar rotation period (45\,d) are highlighted in green and gray, respectively. Shaded vertical bars are plotted at 0.5$\times$ and 2$\times$ the respective signals to emphasize potential aliases. The horizontal dashed lines mark, respectively, the $10\%$ and $1\%$ confidence levels (evaluated with the bootstrap method).
  }
  \label{fig:GLS_activity}
\end{figure*}

\section{Analysis}
\label{sec:analysis}
\subsection{Periodograms of RV and photometric timeseries}
\label{sec:periodogram}

We computed the Generalized Lomb-Scargle (GLS; \citealt{Zechmeister2009}) periodogram of the HARPS-N RVs, using the python package \texttt{astropy} v.5.2.2 \citep{Price2018}. The orbital period of BD+00\,444\,b, about $15.67$ days, is rapidly spotted in the GLS, and it also corresponds to the primary peak of the Maximum Likelihood Periodogram (MLP, \citealt{Stoica1989}), estimated with \texttt{exostriker} v.0.77 \citep{exostriker} (top row of Fig.~\ref{fig:GLS_activity}). Along with the expected signal of the transiting planet, we also found a significant peak (false alarm probability, FAP $\lesssim0.1\%$) at about $97$~d in the GLS, MLP of both RV and RV residuals of the 1-planet model. To verify whether this longer periodicity is related to the stellar activity, we performed the GLS periodograms of several activity indices (bottom panel of Fig.~\ref{fig:GLS_activity}), ranging from the Nyquist frequency of the average time interval to half the full-time coverage. The $97$~d signal is still present in the RV residuals of the 1-planet model (first row of Fig.~\ref{fig:GLS_activity}), but not in the activity markers. Instead, a periodicity of about $45$~d can be appreciated both in the Mount Wilson index, or $S$ index (FAP = 1.3\%), and in the Na\,{\tiny I} doublet (FAP = 1\%), while a peak at about $179$~d is found both in the H$\alpha$ line (FAP = 1\%) and in the area of the cross-correlation function (CCF; FAP = 0.8\%). The area of the CCF is the product between the Full Width Half Maximum (FWHM) and the Contrast of the CCF (for more details, refer to \citealt{Cameron2019}). Here, we utilized the CCF area instead of the FWHM and Contrast because they are strongly anti-correlated, likely due to a sudden change in focus, with a Pearson coefficient of --0.91. 

The GLS of the \emph{TESS} SPOC SAP photometry \citep{Twicken2010,Morris2020} does not reveal any particular periodicity for the short Sector~31. Moreover, we found 6.5 years of photometric data for BD+00\,444 in the All Sky Automated Survey for SuperNovae (ASAS-SN) database, a program that searches for astronomical transients \citep{Kochanek2017}, spanning from September 2017 to March 2024. Computing the GLS periodogram of the Sloan $g$-band time series and after correcting for a linear trend, we retrieved a single significant signal at about $27$~d (Fig.\,\ref{fig:asas}), which might be Moon-related. Overall, we did not find clear indications of stellar activity, although the 45-day signal seems the most plausible as a stellar rotation period. Therefore, we deemed it unlikely that the evident $97$~d signal in the RVs is stellar in origin.

\begin{figure*}
   \centering
   \includegraphics[width=0.8\textwidth]{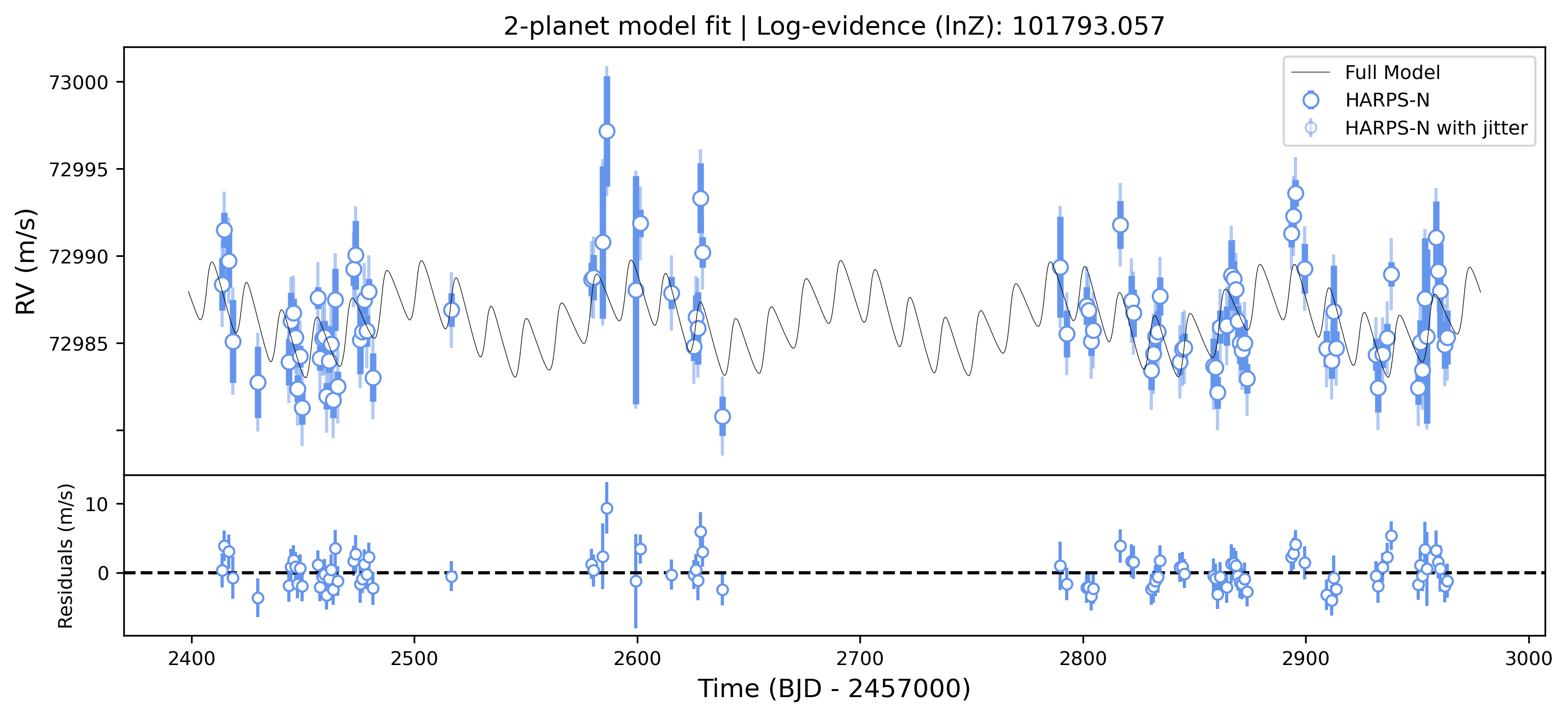}
      \caption{\textit{Top panel}: The HARPS-N RV measurements of BD+00\,444 in blue, along with the preferred model fit in black. \textit{Bottom panel}: the RV residuals over the model fit. The small error bars represent the formal RV uncertainties, while the larger error bars account for uncertainties increased by the best-fit jitter term ($\sigma_w$).}
         \label{fig:RV_2p}
\end{figure*}

\subsection{RV and photometry joint analysis}
\label{sec:joint_analysis}
A joint transit and RV analysis was carried out with \texttt{juliet} following the same approach of \citet{Naponiello2022}. Essentially, in order to reveal the signal of the transiting object from the \emph{TESS} light curve, we first ran the RV and photometry joint analysis with a simple 1-planet model. We used the parameters of the DVR produced by the SPOC pipeline as transit-related priors, both with a fixed null and uniformly-sampled eccentricity via the parametrization $S_1=\sqrt{e}\sin{\omega}$,\, $S_2=\sqrt{e}\cos{\omega}$, which is described by \cite{Eastman2012}. The best 1-planet joint model fit is found with free eccentricity, using a Gaussian prior on the stellar density derived in Sect.~\ref{sec:star}, and it converges to $e=0.30^{+0.05}_{-0.03}$. When the eccentricity is instead fixed to zero, although the values of the impact parameter are similar, the stellar density converges to half the value of its original prior, indicating that the duration of the transit is not compatible with a circular orbit around this star (for more details, refer to \citealt{Dawson2012}). 

Adding a second planet, either with null or free eccentricity, moderately increases the Bayesian evidence, though in the last case its eccentricity is compatible with zero within 1.4$\sigma$ and the evidence is slightly lower. The periodicity of the second signal is well constrained, around 97\,d, even with a large uniform prior of [20, 150] days (Table\,\ref{tab:prior2p}). However, for the circular case, the increase in Bayesian evidence is below the threshold of 5 ($\Delta\ln{\mathcal{Z}}^{2p}_{1p}=3.2$), which is often used to discriminate between positive and strong evidence for more complex models \citep{Kass1995}. Therefore, despite the moderately higher significance, and the flatness of the new residuals compared to the 1-planet model (first two rows of Fig.~\ref{fig:GLS_activity}), here we present this signal as due to planet~candidate BD+00\,444\,c, with its properties detailed in Table~\ref{tab:pparameters}. From this point onward, we will consider the parameters obtained from the 2-planet model as more likely to be accurate, especially because the parameters of the transiting planet~BD+00\,444\,b (also listed in Table~\ref{tab:pparameters}) are fully consistent between the models (e.g. the bulk density is consistent well within 1$\sigma$).

For comparison, we employed Gaussian processes (GPs) to model the second signal as stellar activity. However, the hyper-parameter posteriors of the 1-planet + GP model are not well constrained and, specifically, the stellar rotation period does not converge when using the quasi-periodic kernel. The lack of convergence of the rotation period, despite having the same uniform prior as the one used for the planetary period of candidate BD+00\,444\,c in the 2-planet model, indicates that a simple Keplerian model is a better fit to the data. This is consistent with the estimated $\log R^{\prime}_{\rm HK}$ of Table\,\ref{tab:star}. Similarly, when the GP quasi-periodic kernel is employed along with the 2-planet model, the parameters of the outer candidate are still well retrieved, albeit with higher uncertainty. For this reason, and because we did not have any indications of strong stellar activity, we decided not to include GPs in the final model. In particular, we deemed unnecessary the use of multi-dimensional GPs, as the RV residuals of the 1-planet model are not correlated with any activity index (e.g., the residuals show minimal correlation only with H$\alpha$, yielding a Pearson correlation coefficient of 0.33). 

\begin{figure}
  \centering
  \begin{subfigure}{\linewidth} 
   \centering
   \includegraphics[width=1\textwidth]{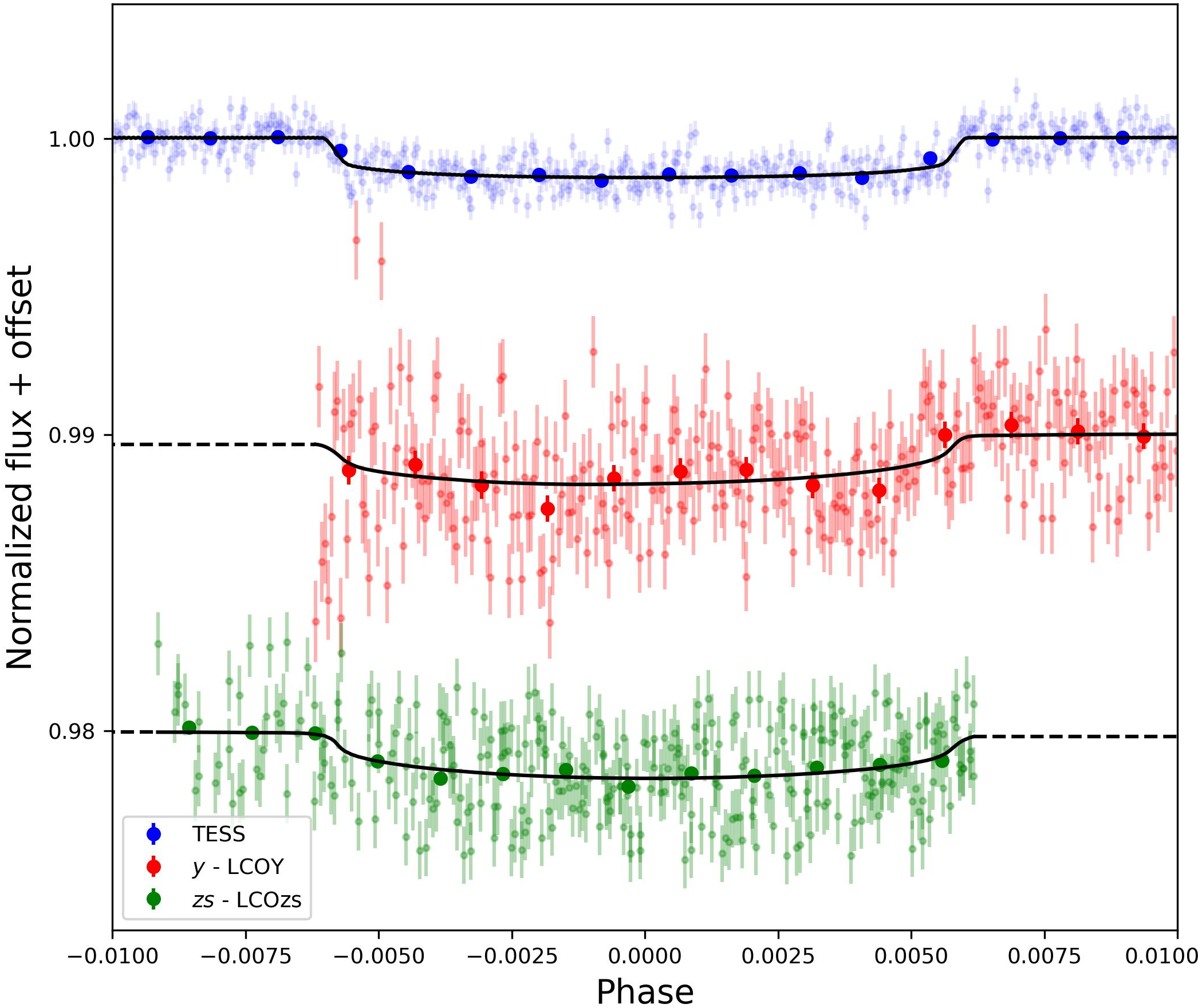}
  \end{subfigure}
  \begin{subfigure}{\linewidth}
    \centering
    \includegraphics[width=1\textwidth]{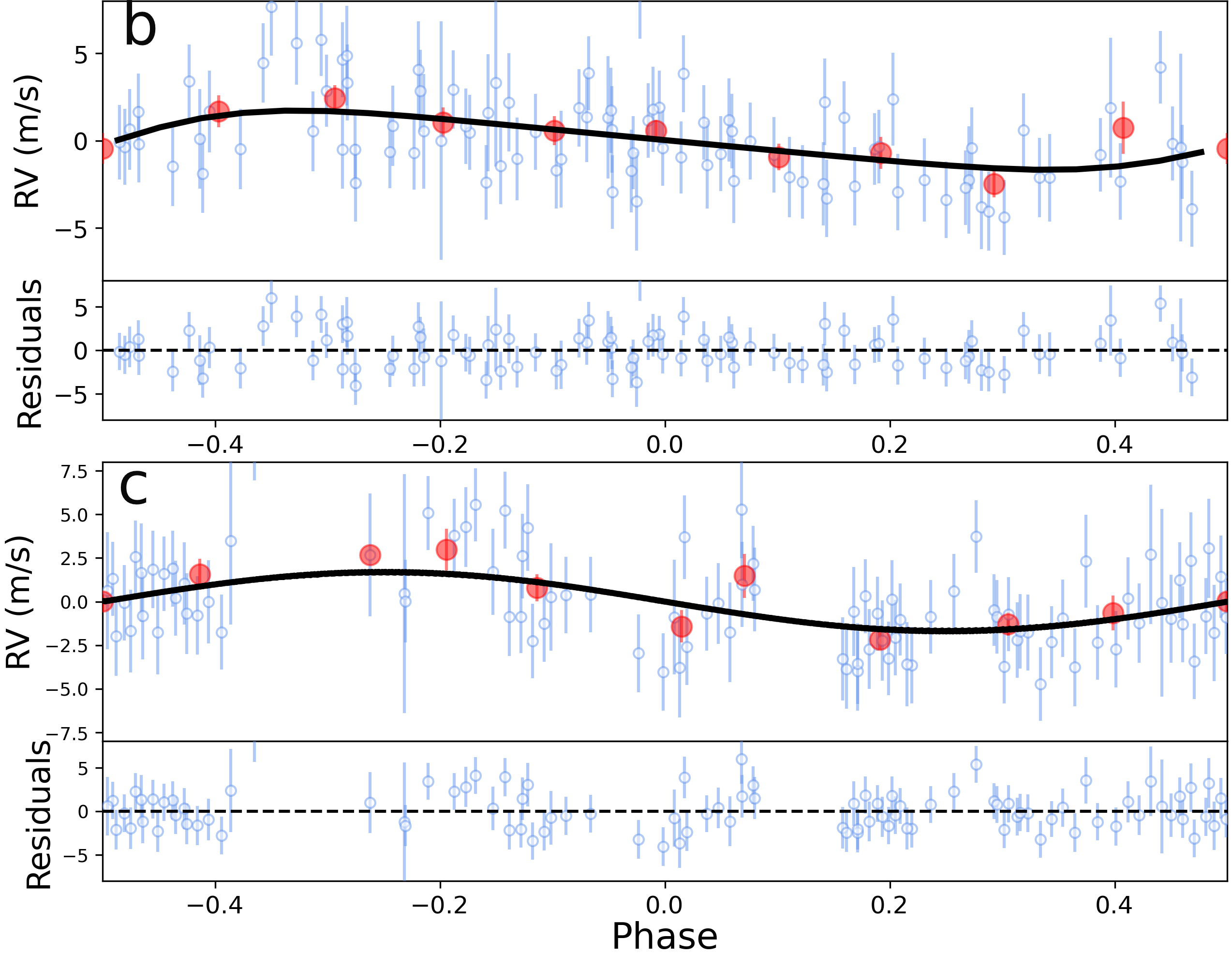}
  \end{subfigure}
    \caption{{\it Top panel:} Global fit result for \emph{TESS} and the ground-based transits. Both LCOGT light curves are shifted on the $y$-axis for clarity, and their respective filter band are indicated in the legend, while the superimposed points represent $\sim$ 30-minute bins. {\it Bottom panels:} Phase-folded HARPS-N RVs to the period of planet~b and candidate~c, along with their residuals. The red circles represent the average value of $\sim10$ phased RV data points at a time.}
    \label{fig:phase2p}
\end{figure}

Finally, the RVs are shown in Fig.~\ref{fig:RV_2p} along with the preferred global model and its residuals, while in Fig.~\ref{fig:phase2p} the phase-folded \emph{TESS} transits and the phase-folded RVs are plotted along with the two transits observed by LCOGT.
\begin{table}
\centering
\caption{Planet parameters.}\label{tab:pparameters}
\renewcommand{\arraystretch}{1.2}
\resizebox{\hsize}{!}{\begin{tabular}{lccc}
    \hline\hline
     Parameters & BD+00\,444\,b & BD+00\,444\,b & BD+00\,444\,c \\
     & (1p model) & (2p model) & (2p model) \\ [2pt]
\hline \\[-6pt]%
\multicolumn{1}{l}{\large{Transit and orbital}} \\[2pt]
$K$ (m\,s$^{-1}$)\dotfill & $2.23^{+0.37}_{-0.38}$ & $1.74^{+0.40}_{-0.39}$ & $1.73^{+0.34}_{-0.37}$ \\
$P_{\rm orb}$ (d)\dotfill & $15.66858^{+0.00011}_{-0.00015}$ & $15.66858\pm0.00010$ & $96.6\pm1.4$ \\
$T_{\rm 0}$ (BJD - $2\,459\,000$)\dotfill & $148.09923\pm0.00060$ & $148.09939\pm0.00051$ & $235.2^{+6.4}_{-6.2}$ \\
$T_{\rm 14}$ (h)\dotfill & $4.548^{+0.054}_{-0.046}$ & $4.548^{+0.053}_{-0.047}$ & ... \\
$R_{\rm p}/R_{\star}$\dotfill & $0.03432\pm0.00069$ & $0.03436\pm0.00062$ & ... \\ 
$b$\dotfill & $0.25\pm0.15$ & $0.22\pm0.15$ & ...\\ 
$i$ (deg)\dotfill & $89.69^{+0.19}_{-0.17}$ &  $89.72^{+0.17}_{-0.18}$ & ... \\ 
$a/R_{\star}$\dotfill & $35.8^{+0.76}_{-0.79}$ & $35.8^{+0.74}_{-0.76}$ & $120\pm3$ \\
$q_1$\dotfill & $0.25^{+0.18}_{-0.13}$ & $0.26^{+0.21}_{-0.14}$ & ... \\
$q_2$\dotfill & $0.34^{+0.25}_{-0.21}$ & $0.30^{+0.22}_{-0.18}$ & ... \\
$\sqrt{e}\sin\omega$\dotfill & $-0.55^{+0.03}_{-0.04}$ & $-0.54^{+0.03}_{-0.04}$ & ... \\
$\sqrt{e}\cos\omega$\dotfill & $0.02\pm0.11$ & $0.04\pm0.11$ & ... \\[2pt]
\multicolumn{1}{l}{\large{Derived}} \\[2pt]
$M_{\rm p}$ ($M_{\oplus}$)\dotfill & $6.1^{+1.0}_{-1.1}$ & $4.8\pm1.1$ & ... \\
$M_{\rm p}\sin{i}$ ($M_{\oplus}$)\dotfill & ... & ...  & $9.3^{+1.8}_{-2.0}$ \\[2pt]
$R_{\rm p}$ ($R_{\oplus}$)\dotfill & $2.363\pm0.066$ & $2.363\pm0.066$ & ...\\
$\rho_{\rm p}$ (g\,cm$^{-3}$)\dotfill & $2.56^{+0.48}_{-0.46}$ & $2.00^{+0.49}_{-0.45}$ & ... \\ [2pt]
$\log{g_{p}}$ (cgs)\dotfill & $10.8\pm1.9$ & $8.4^{+2.0}_{-1.9}$ & ... \\
$a$ (au)\dotfill & $0.1051\pm0.0026$ & $0.1050\pm0.0025$ & $0.3529\pm0.0093$ \\ 
$T_{\rm eq}^{(\ddagger)}$ (K)\dotfill & $519\pm6$ & $519\pm6$ & $283\pm4$ \\
$e$\dotfill & $0.308^{+0.047}_{-0.035}$ & $0.301^{+0.046}_{-0.034}$ & <0.50 \\ [2pt]
$\omega$ (deg)\dotfill & $-88^{+9}_{-11}$ & $-86\pm11$ & unconstrained \\ 
TSM$^{(*)}$\dotfill & $124^{+28}_{-19}$ & $159^{+46}_{-31}$ & ... \\ [2pt]
$u_1$ (TESS)\dotfill & $0.34^{+0.15}_{-0.14}$ & $0.31^{+0.14}_{-0.13}$ & ... \\
$u_2$ (TESS)\dotfill & $0.15^{+0.14}_{-0.12}$ & $0.19^{+0.13}_{-0.12}$ & ... \\
$u_1$ (LCO$_Y$)\dotfill & $0.72^{+0.30}_{-0.28}$ & $0.78^{+0.30}_{-0.27}$ & ... \\
$u_2$ (LCO$_Y$)\dotfill & $0.06^{+0.28}_{-0.27}$ & $-0.02^{+0.26}_{-0.26}$ & ... \\
$u_1$ (LCO$_{zs}$)\dotfill & $0.70^{+0.28}_{-0.25}$ & $0.79^{+0.28}_{-0.25}$ & ... \\
$u_2$ (LCO$_{zs}$)\dotfill & $0.00^{+0.23}_{-0.24}$ & $-0.11^{+0.23}_{-0.24}$ & ... \\
\multicolumn{1}{l}{\large{Instrumental}} \\[2pt]
$\sigma_{\textsf{w,TESS}}$ (ppt)\dotfill & $0.000070^{+0.000018}_{-0.000012}$ & $0.000070^{+0.000021}_{-0.000012}$ & ... \\
$\rho_{\textsf{TESS}}$ (d)\dotfill & $0.72^{+0.48}_{-0.28}$ & $0.67^{+0.46}_{-0.25}$ & ... \\
$\overline{\mu}_{\textsf{HARPS-N}}$ (m\,s$^{-1}$)\dotfill & $72986.06^{+0.24}_{-0.25}$ & $72986.35^{+0.23}_{-0.24}$ & ... \\
$\sigma_{\textsf{w,HARPS-N}}$ (m\,s$^{-1}$)\dotfill & $2.25\pm0.20$ & $1.96^{+0.24}_{-0.19}$ & ... \\

    \bottomrule
\end{tabular}
}
\tablefoot{Best-fit median values, with upper and lower 68\% credibility bands as errors, of the fitted and derived parameters for BD+00\,444\,b and BD+00\,444\,c, as extracted from the posterior distribution of the relative models (refer also to Table~\ref{tab:prior2p} and Fig.~\ref{fig:corner2p}). $^{(\ddagger)}$ This is the equilibrium temperature for a zero Bond albedo and uniform heat redistribution to the night side. The eccentricity upper limit is constrained at the confidence level of 1$\sigma$. $^{(*)}$ Transmission spectroscopy metric (TSM; \citealt{Kempton2018}).}
\end{table}

\subsection{HARPS-N detection sensitivity}
\label{sec:detection}
\begin{figure}
    \centering
    \includegraphics[width=0.49\textwidth]{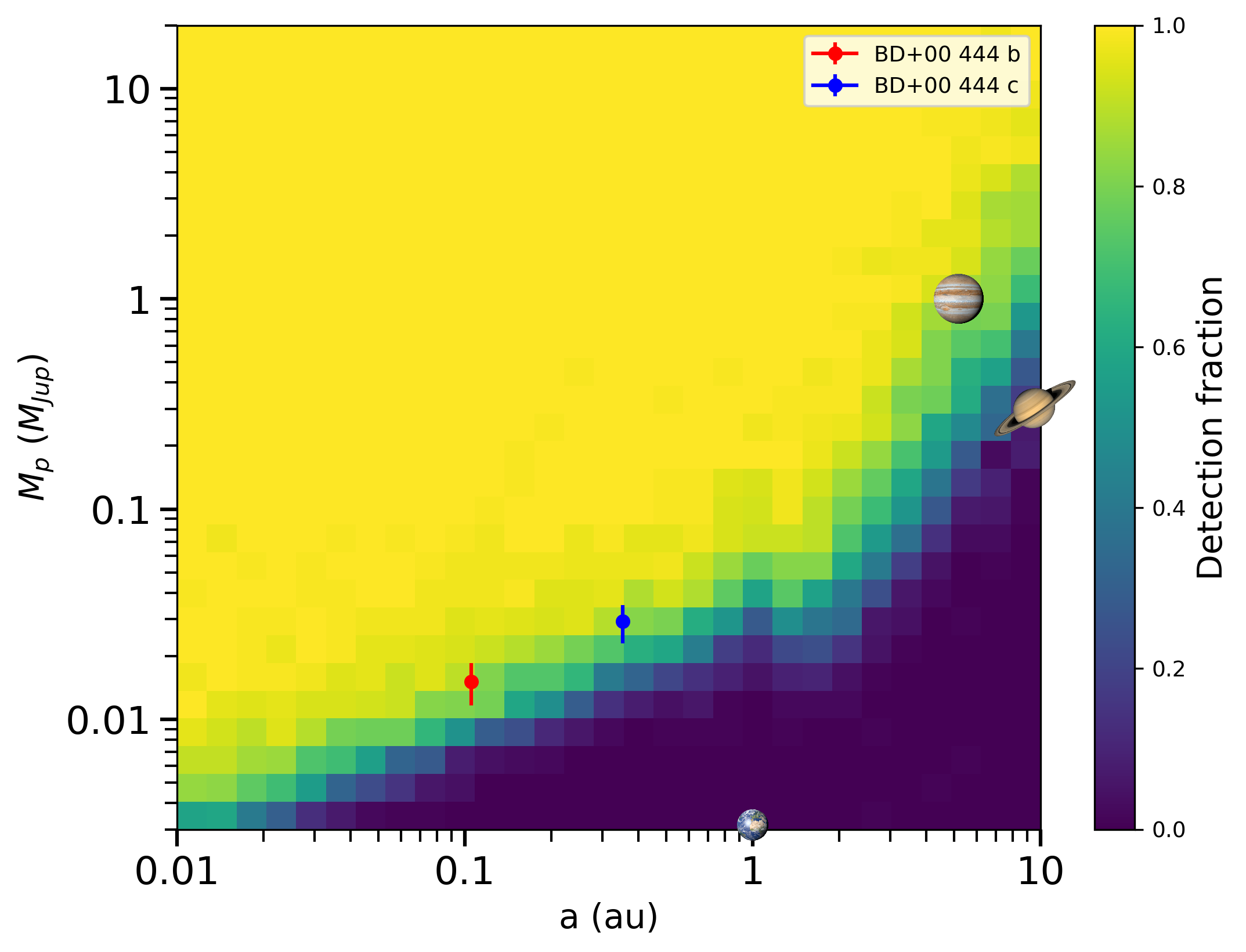}
    \caption{HARPS-N RV detection map for BD+00\,444. The color scale expresses the detection fraction (i.e. the detection probability), while the circles mark the position of BD+00\,444\,b (in red) and candidate BD+00\,444\,c (in blue), for which we use $M_p\sin{i}$. Jupiter, Saturn and the Earth are shown for comparison.}
    \label{fig:completeness}
\end{figure}

We estimated the completeness of the HARPS-N RV time series by performing injection-recovery simulations, in which synthetic RVs with planetary signals were injected at the times of our observations, using HARPS-N error bars and the estimated stellar jitter included in Table\,\ref{tab:prior2p}. We simulated the signals of additional companions across a logarithmic $30 \times 30$ grid in planetary mass ($M_{\rm p}$) and semi-major axis ($a$), covering the ranges 0.01--20\,$M_{\rm Jup}$ (or Jupiter masses) and 0.01--10\,au. As in \citet{Bonomo2023}, for each location of the grid we generated 100 synthetic planetary signals, drawing $a$ and $M_{\rm p}$ from a log-uniform distribution inside the cell, $T_0$ from a uniform distribution in $[0,P]$, the orbital inclination $i$ from a uniform $\cos{i}$ distribution in $[0,1]$, $\omega$ from a uniform distribution in $[0,2\pi]$, and $e$ from a beta distribution in $[0.0, 0.8]$ \citep{Kipping2013}. We fitted the signals by employing either Keplerian orbits or linear and quadratic trends, in order to take into account long-period signals, which would not be correctly identified as Keplerian due to the short time span of the RV observations (550\,d). We then adopted the Bayesian information criterion (BIC) to compare the fitted planetary model with a constant model and considered the planetary signal significantly detected only when $\Delta \text{BIC} > 10$ in favor of the planet-induced one. The detection fraction was finally computed as the portion of detected signals for each grid element, as illustrated by Fig.\,\ref{fig:completeness}.

\section{Discussion}
\label{sec:discussion}
\subsection{The non-transiting candidate BD+00\,444\,c}
\label{sec:candidate}

\begin{figure}
\centering
\includegraphics[width=0.5\textwidth]{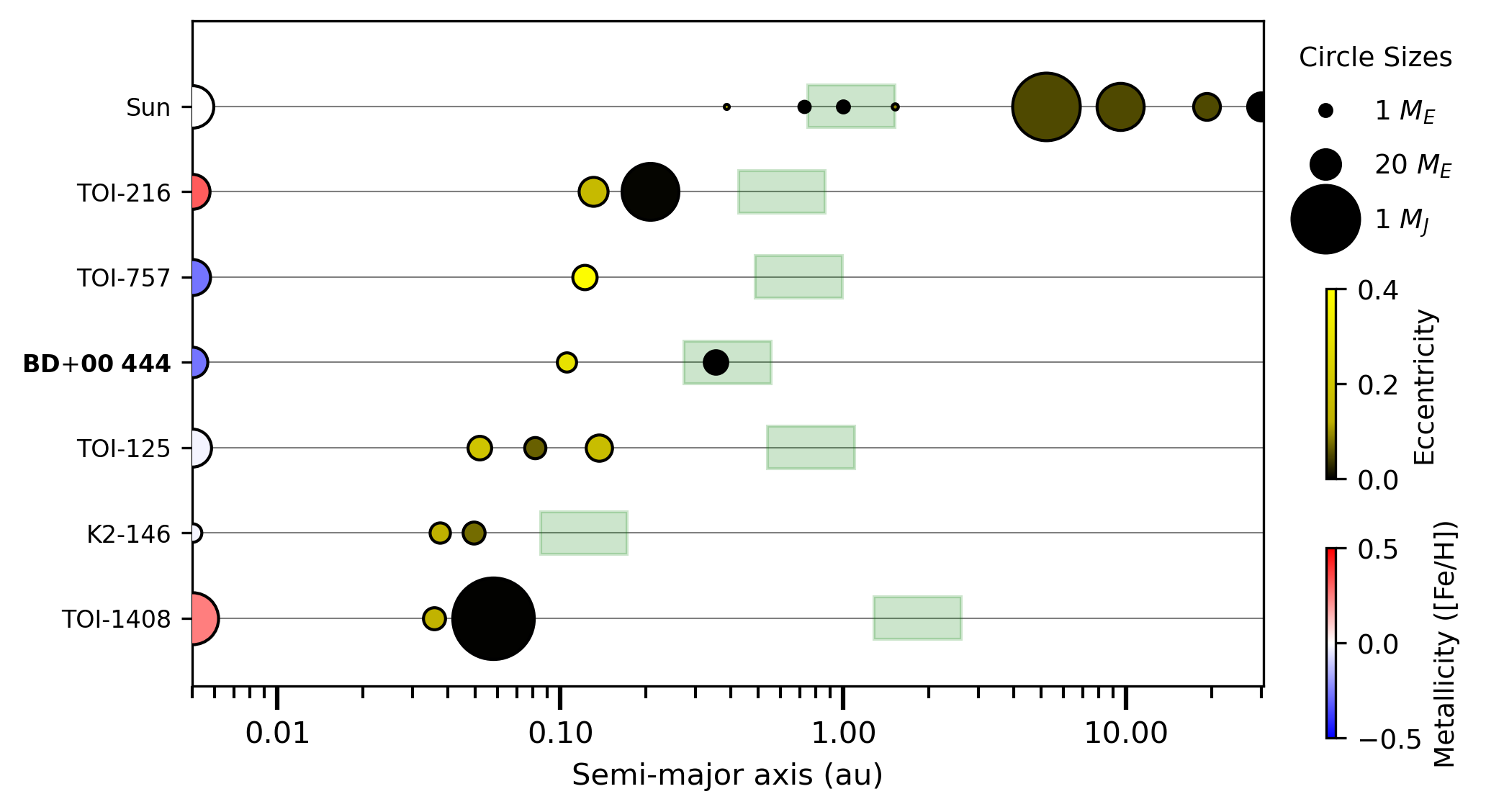}
\caption{System architectures featuring inner transiting planets with masses below $20M_{\oplus}$ (represented with different circle sizes) and eccentricities greater than 0.10 (represented with yellow color), determined with at least 5$\sigma$ confidence. The shaded green rectangles represent the respective habitable zones, while the host metallicity and size are illustrated by color and circle size, respectively. The Solar System is included for comparison.}
\label{fig:peas}
\end{figure}

We infer that the true mass of BD+00\,444\,c should lie within a factor of 2 of the minimum mass ($M_{\rm c}\sin{i}=9.3^{+1.8}_{-2.0}M_{\oplus}$). This is because the probability that its orbital inclination lies between 30 deg and 90 deg is $\sim$87\% \citep{fischer2014}, without even taking into account the existence of the close-in transiting planet, with an inclination of $\sim$90 deg. The predicted time of transit of BD+00\,444\,c is outside the region observed by \emph{TESS}, and its host star will not be observed in the upcoming \emph{TESS} sectors (at least up to early 2026). However, with an estimated posterior transit probability of $\sim$1\%, adopting the formulation given by \citet{Stevens2013} and assuming $e=0$, it is possible that the candidate is not transiting at all. Therefore, we expect that the planet will necessitate more RV measurements to be secured with higher statistical evidence. 

Interestingly, BD+00\,444\,c has an equilibrium temperature of $283\pm4$ K (assuming zero Bond albedo and uniform heat redistribution to the night side) that would place it within the habitable zone (Fig.\,\ref{fig:peas}). Despite its location, this planet likely has a Neptune-type composition, which would render it inhospitable for life as we know it. Nevertheless, a moon orbiting this planet might harbor conditions suitable for the development of life (see, e.g., \citealt{Martinez2019}).

\subsection{Composition of BD+00\,444\,b}
\label{sec:composition}
We employed an advanced interior model to study BD+00\,444\,b, where water can be present in the core, the mantle, and surface depending on the specific thermal state of the planet. The model is based on that of \citet{dorn_generalized_2017} with recent adaptations as presented by \citet{luo_majority_2024}. For this application, we focused on interiors with Earth-like rocky interiors with H$_2$-He atmospheres, water steam atmospheres, and atmospheres containing a mixture of  water and H$_2$-He. In the last case, we considered two different envelope structures: a homogeneously mixed atmosphere consisting of H$_2$-He and water, and a differentiated structure where the water and H$_2$-He layers are separated, with the pure water layer below the pure H$_2$-He layer. Although the homogeneous envelope is more realistic, we added the layered case for comparison.

We considered a core made of Fe with the light alloy elements H and O. For solid Fe, we used the equation of state for hexagonal close packed iron \citep{Hakim_new_2018,Miozzi_new_2020}. For liquid iron and iron alloys, we use the equation of state by \citet{luo_majority_2024}. The core thermal profile is assumed to be adiabatic. At the core-mantle boundary, there is a temperature jump as the core can be hotter than the mantle due to the residual heat released during core formation following \citet{stixrude_melting_2014}.

The mantle is assumed to be made up of three major constituents, i.e., MgO, SiO$_2$, and FeO. For the solid mantle, we used the thermodynamical model \verb|Perple_X| \citep{connolly_geodynamic_2009} to compute stable mineralogy and density for a given composition, pressure, and temperature, employing the database of \citet{Stixrude_thermal_2022}. For pressures higher than $\sim125$~GPa, we defined stable minerals a priori and used their respective equation of states from various sources \citep{Hemley_constraints_1992, Fischer_equation_2011, Faik_equation_2018, Musella_physical_2019}. For the liquid mantle, we calculated its density assuming an ideal mixture of the main components (Mg$_2$SiO$_4$, SiO$_2$, and FeO)  \citep{melosh_hydrocode_2007,Faik_equation_2018,ichikawa_ab_2020,stewart_shock_2020} and added them using the additive volume law. We used Mg$_2$SiO$_4$ instead of MgO since the data for forsterite was recently updated for the high-pressure temperature regime \citep{stewart_shock_2020}, which is not available for MgO to our knowledge. The mantle is assumed to be fully adiabatic.

Water can be present in mantle melts, while solid mantle is assumed to be dry. The addition of water reduces the density, for which we follow \citet{Bajgain_structure_2015}. For small water mass fractions, this reduction is nearly independent of pressure and temperature. The melting curve of mantle material is calculated for dry and pure MgSiO$_3$. The addition of water \citep{Katz_new_2003} and iron \citep{dorn_outgassing_2018} can lower the melting temperatures. Water added to the core also lowers its melting temperature, for which we followed \citet{luo_majority_2024}. Possible water in the core may be present in both liquid and solid phases.
The partitioning between mantle melts and the water layer is determined by Henry’s law, for which we used the fitted solubility function of \citet{Dorn_hidden_2021}. For the partitioning of water between iron and silicates, we followed \citet{luo_majority_2024}. For the equilibration pressure of water to partition between iron and silicates, we used half of the core-mantle boundary pressure, which is within typically discussed values for Earth (0.3--0.6). Varying this pressure would introduce overall small changes in the distribution of water.

On BD+00\,444\,b water can be in steam or supercritical phase, for which we use the AQUA \citep{Haldemann_aqua_2020} compilation of equation of states. For pressures below 0.1 bar, we assumed an isothermal profile and then switched to an adiabatic profile. Whenever we considered H$_2$-He, we followed the model of \citet{guillot_radiative_2010} for an irradiated atmosphere and an underlying gaseous envelope. If H$_2$O and H$_2$-He are fully mixed, we adjust the metallicity $Z$ of the envelope according to 
\begin{equation}
    Z = \frac{m_{\rm H_2O}}{m_{\rm H_2O}+m_{\rm H_2-He}},
\end{equation}
where $m_{\rm H_2O}$ and $m_{\rm H_2/He}$ are the total H$_2$O and H$_2$-He mass of the envelope respectively. Further details on the structure models can be found in \citet{dorn_generalized_2017}. 

\begin{figure*}[!t]
    \centering
    \includegraphics[angle=90, width=0.50\textwidth]{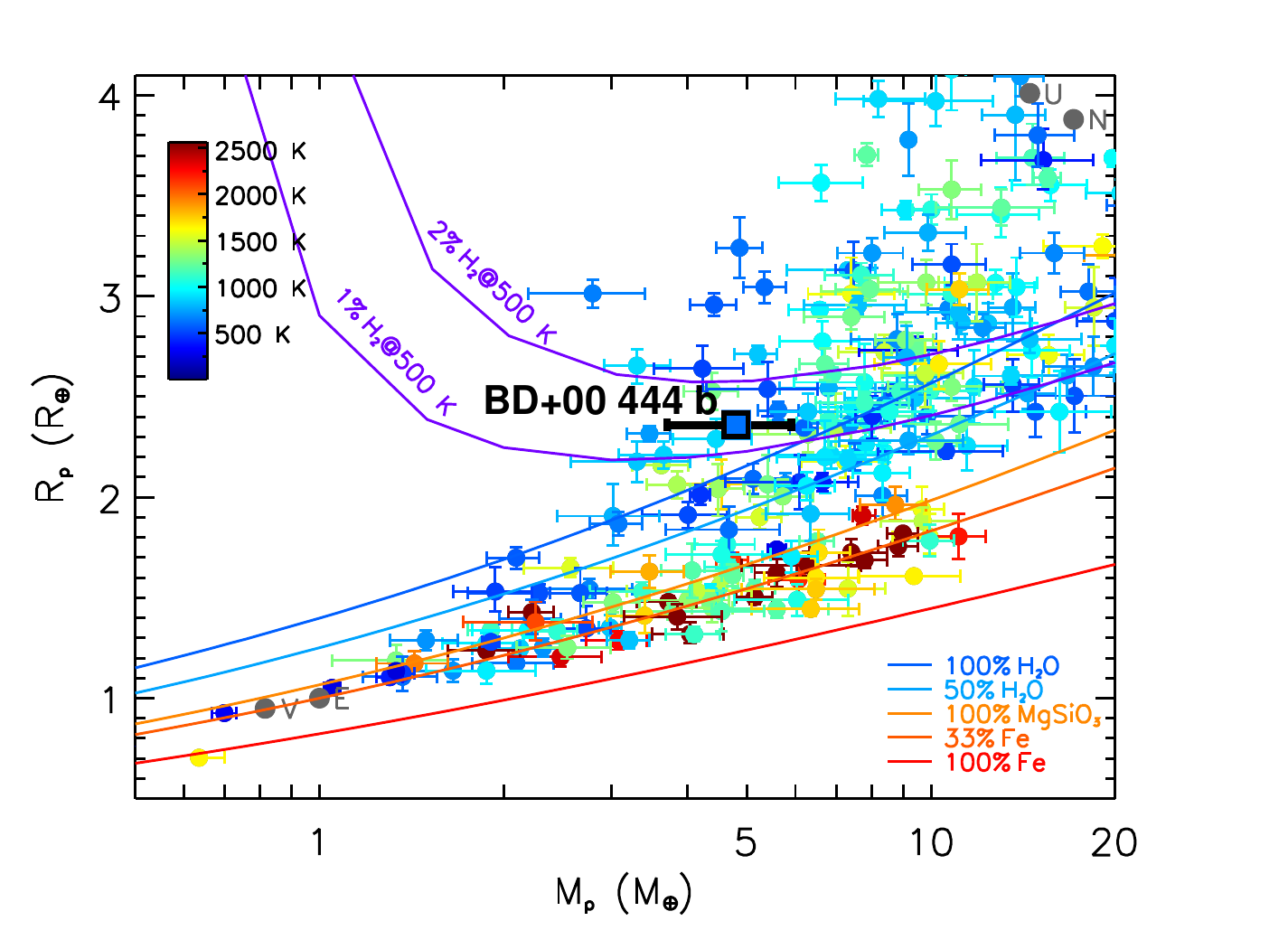}\hspace{-6mm}
    \includegraphics[angle=90, width=0.50\textwidth]{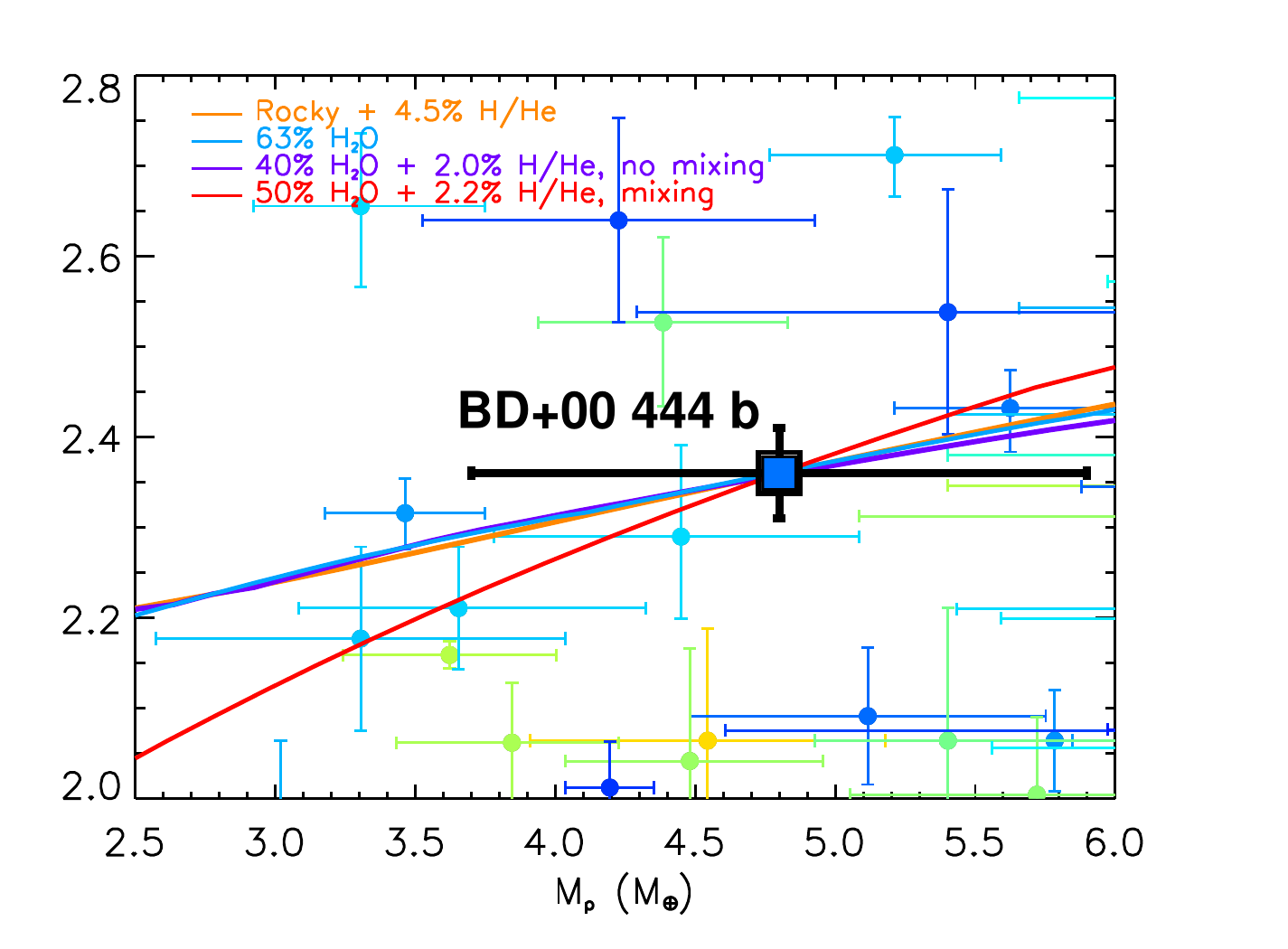}
    \caption{\textbf{Left panel:} Planetary masses and radii of the known transiting exoplanets. The color of the planets corresponds to their equilibrium temperatures. The iso-composition curves for differentiated planets \citep{Zeng2013} are described in the legend. \textbf{Right panel:} Zoom-in of the same diagram in mass linear scale, 
    displaying the different compositions that best match the mass and radius of BD+00\,444\,b with both differentiated and miscible interiors  (Sect.~\ref{sec:composition}).}
    \label{fig:mass-radius}
\end{figure*}


It is possible to match the bulk properties of BD+00\,444\,b with all four structure models. In the left panel of Fig.~\ref{fig:mass-radius}, we show mass-radius relations between known transiting exoplanets\footnote{Values taken from the Transiting Extrasolar planet~catalog, TEPCat, which is available at \url{www.astro.keele.ac.uk/jkt/tepcat/}; \citealt{Southworth2010,Southworth2011}} and classical layered structure models by \citet{Zeng2013}, while, in the right panel, we display the resulting relations from our best-fit parameters. For a pure water-world planet without any H$_2$-He, the bulk properties of BD+00\,444\,b are consistent with a water mass fraction of 60\%--65\%. This is slightly higher than the commonly used upper limit of 50\% \citep{luque_density_2022}. However, the atmosphere contains only 6.2--7.5\% of the total water content. The majority of the water is stored in the interior of the planet. Considering the other extreme case, where no water is present on the planet, the bulk properties are consistent with a H$_2$-He mass fraction of $4.5\pm0.5\%$. More realistically, the atmosphere of the planet~contains both water and H$_2$-He. If the two components are separate, the water mass fraction is $40\%$ with $2.0\pm0.4\%$ H$_2$-He. If, on the other hand, the H$_2$O and H$_2$-He are fully mixed, the bulk properties are consistent with a water fraction of $50\%$ and a H$_2$-He fraction of $2.2\pm0.5\%$.

\subsection{Formation and evolution}
\label{sec:formation}
We investigated the possible formation history of BD+00\,444's two planets following the approaches described by \citet{Mantovan2024} and \citet{Damasso2024}. We used our Monte Carlo version of the {\tt GroMiT} (planetary GROwth and MIgration Tracks) code \citep{Polychroni2023} to simulate the possible formation tracks of planets b and c in the framework of the pebble accretion scenario. In particular, we employed the treatments for the growth and migration of pebble-accreting and gas-accreting planets from \citet{Johansen2019} and \citet{Tanaka2020} and the scaling law for the pebble isolation mass from \citet{Bitsch2018}.

The planet formation model builds on the description of the viscously evolving circumstellar disk in which the planets are embedded \citep{Johansen2019,Armitage2020} complemented by the prescriptions for the characterization of its thermal profile due to the interplay between viscous heating and stellar irradiation from \citet{Ida2016}, and of its solid-to-gas ratio profile from \citet{Turrini2023}. The stellar luminosity during the pre-main sequence phase was set to 0.85 $L_\odot$ based on the stellar evolutionary models from \citet{Baraffe2015} for stars with the mass of BD+00\,444 at the age of 1\,Myr. 

\begin{figure}
    \includegraphics[width=0.5\textwidth]{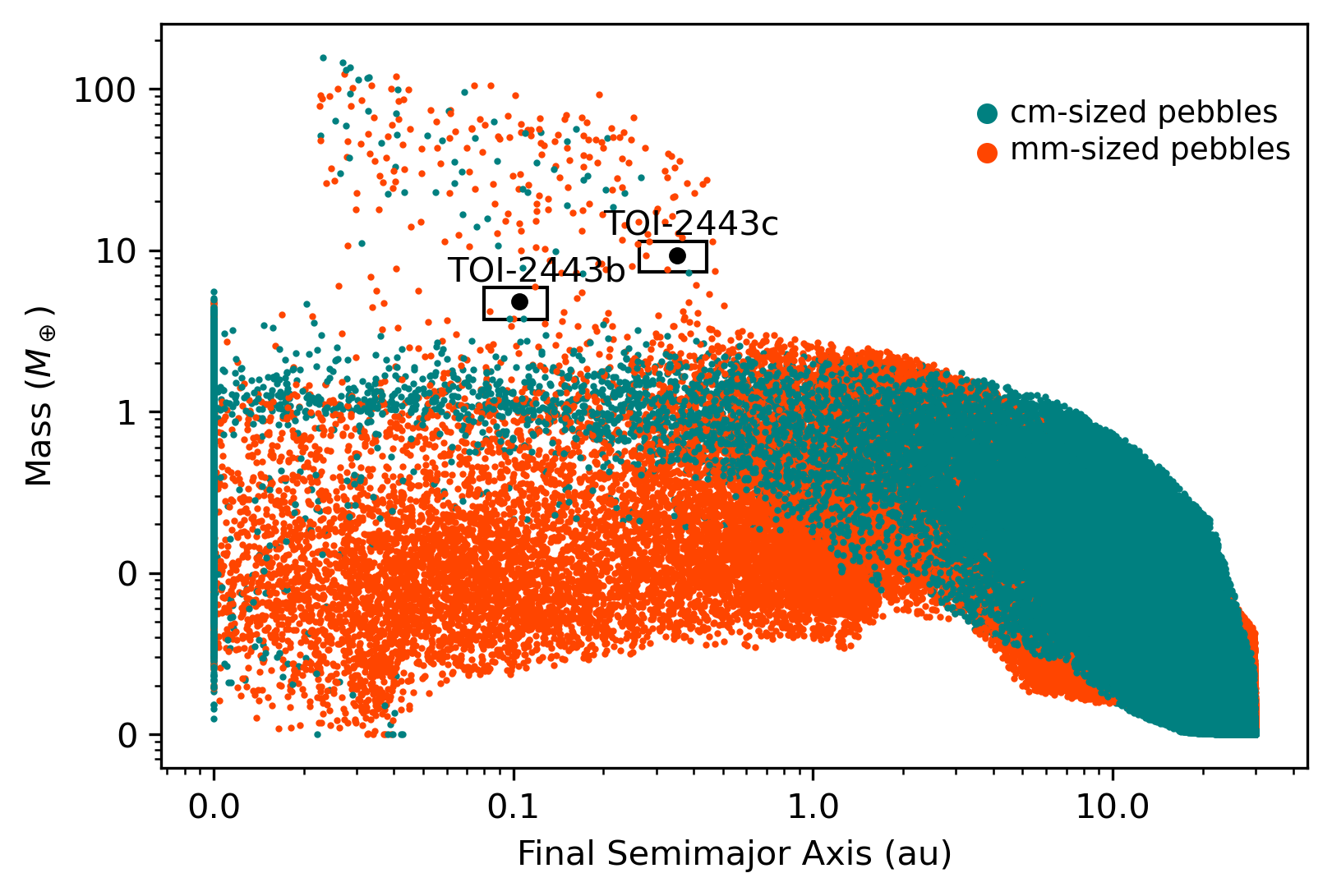}
    \caption{Synthetic populations of planets resulting from the two Monte Carlo runs of 2$\times$10$^5$ extractions each with {\tt GroMiT}. The plot shows the final masses and orbital periods of the simulated growth tracks. The vermilion symbols indicate those planets formed in disks dominated by mm-sized pebbles, while with cactus green we denote cm-sized pebble-dominated disks. Planets b and c are indicated using two larger cyan circles and the boxes around them highlight the region of parameter space populated by the successful solutions.}
    \label{fig:popsyn}
\end{figure}

We ran two sets of simulations considering a millimeter-sized (mm-sized) pebble-dominated disk and a centimeter-sized (cm-sized) one, each one simulating the formation of 2$\times$10$^5$ individual planets by randomly varying the formation time and the initial semi-major axis of their planetary seeds, as well as the disk viscosity coefficient, $\alpha$. We kept the rest of the physical characteristics of the disk constant for all runs; see Table \ref{tab:popsythesis} for more details. In Fig. \ref{fig:popsyn} we plot the complete results of our simulations, as well as the locations where planet~b and candidate c fall in.

\begin{figure}
    \includegraphics[width=0.5\textwidth]{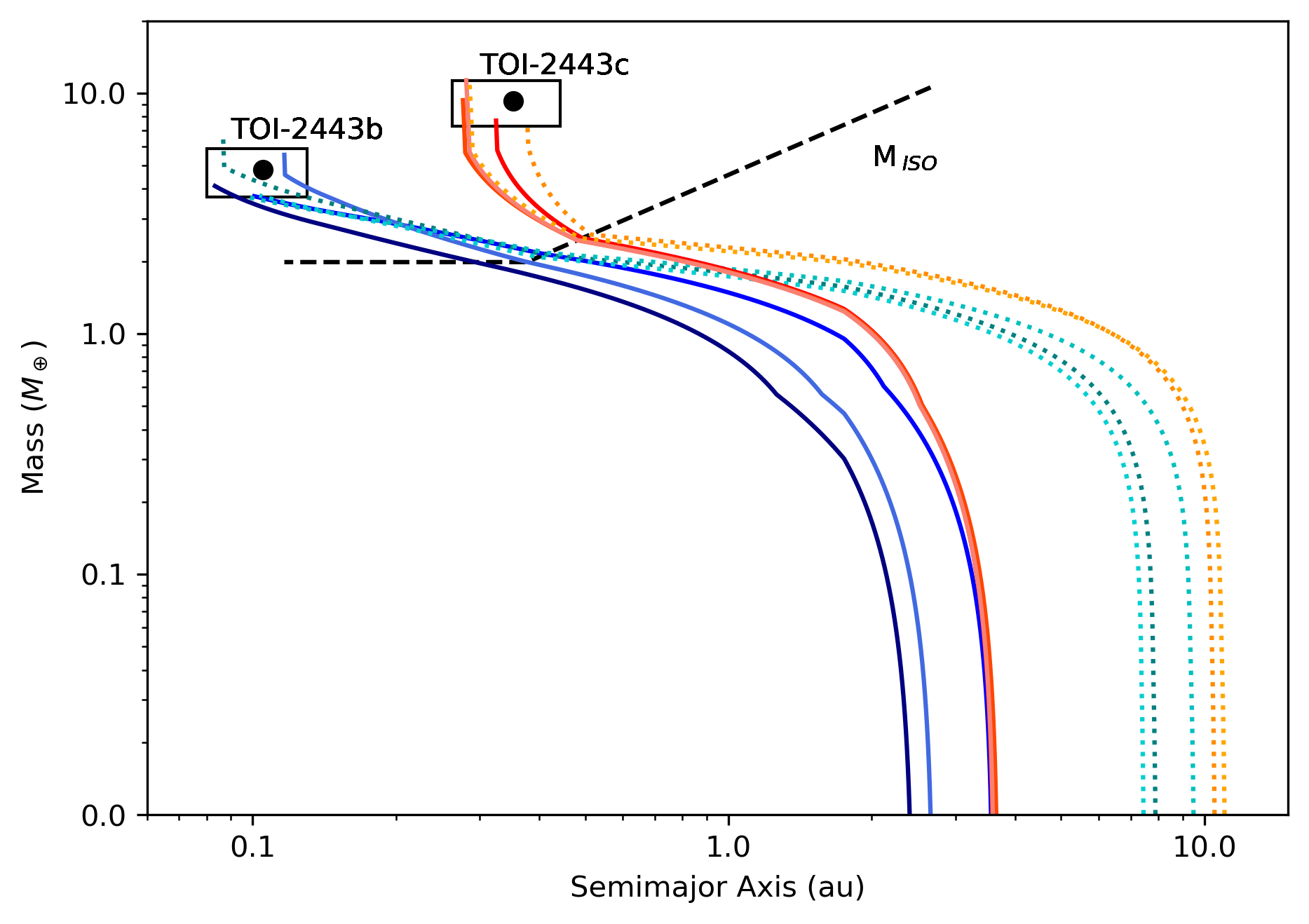}
    \caption{Planetary growth tracks that satisfy the selection conditions associated with the black boxes in Fig. \ref{fig:popsyn}. The growth tracks are projected in the semi-major axis--planetary mass space. The blue tracks denote the successful solutions for planet~b while the red ones are those of planet~c. The solid lines indicate the planets that formed in an mm-sized pebble-dominated disk, while the dotted ones formed in a cm-sized one. $M_{iso}$ denotes the pebble isolation mass above which no pebbles are able to accrete onto the forming planet \citep{Lambrechts2014}.}
    \label{fig:tracks}
\end{figure}

We considered as successful those solutions where the resulting objects have (i) final mass within 1$\sigma$ of the reported mass of each one (in the case of candidate c we focus on its minimum mass) and (ii) final semi-major axis within half and twice the nominal semi-major axis of each planet, to account for the uncertainty of the planetary migration rates \citep[e.g.][]{Pirani2019}. These solutions are plotted in Fig. \ref{fig:tracks}. We found that the modeled system can create a few massive planets, in line with its reported low metallicity. All successful synthetic planets emerge from planetary seeds that formed within the first 1 Myr of the disk's lifetime, with the planet~b's solutions starting their growth on average slightly later than the planet~c's.

In the cm-sized pebble-dominated disk, the successful solutions of both planets start forming beyond 7\,au (on average at 8\,au for planet~b and at 11\,au for planet~c) and need their planetary seed to appear very early in the disk's lifetime (within the first 0.4 Myr) to successfully gather the required mass and migrate inwards. In the mm-sized pebble-dominated disk, both planets start their formation within 4\,au (on average 3.0\,au and 3.6\,au respectively) and their seeds appear later, up to 0.7 Myr. Also in this case candidate planet~c generally starts its growth a bit earlier than planet~b. 

Regardless of the specific planet and pebble size, all successful solutions complete the growth of their core at about the same time within a similar turbulent environment ($\alpha$=0.0022--0.0028), forming similar cores of 2.0--2.5\,$M_\oplus$. While we cannot differentiate between the cases of mm-sized and cm-sized pebbles based solely on the physical characteristics of the two planets, there are some conclusions that we can reach. Candidate planet~c needs to start its formation earlier and further out than planet~b, in order to reach its reported minimum mass in all scenarios. However, given that the two planetary cores complete their growth pretty much by the same time (i.e. the growth of planet~c is slower than that of planet~b) it is unlikely that the outer object experienced close encounters with the inner one during the formation stages.

To gain further insight into the possible formation environment of BD+00\,444's planetary system, we compared the core masses emerging from our planet formation model for planet~b with those allowed by the mass-radius relationships for the cores and the gaseous envelopes from \citet{Fortney2007} and \citet{Lopez2014}. Based on the results on the planetary compositions compatible with BD+00\,444\,b's characteristics discussed in Sect. \ref{sec:composition} and the formation separations discussed above, we explored three core compositions. The first one is an ice-rich core composition described by Eq. 7 from \citet{Fortney2007} where we set the water ice mass fraction in the core to 0.33 based on the mass balance between rocks and ices from \citet{Turrini2021}. The second one is a rocky composition based on Eq. 8 from \citet{Fortney2007} where the rock mass fraction in the core is set to 0.66 to assume an Earth-like composition. The last one is the simplified treatment for a rock-iron core from Eq. 2 of \citet{Lopez2014}.

For the envelope, we considered both the cases of solar metallicity and enhanced (50$\times$ solar) metallicity from \citet{Lopez2014}. Given the mature age of BD+00\,444's star, however, the differences in the envelope radii associated with these two compositions are significantly smaller than those due to the other parameters that we sampled. As in \citet{Damasso2024}, for each combination of core-envelope compositions, we performed $10{^6}$ Monte Carlo extractions where we randomly sampled the core mass fraction, stellar age, and planetary mass. The core mass fraction is extracted from a uniform distribution between 0 and 1, while the stellar age and planetary mass are extracted from normal distributions truncated to zero whose widths are set by the uncertainties on the two parameters.

\begin{figure}
    \includegraphics[width=0.5\textwidth]{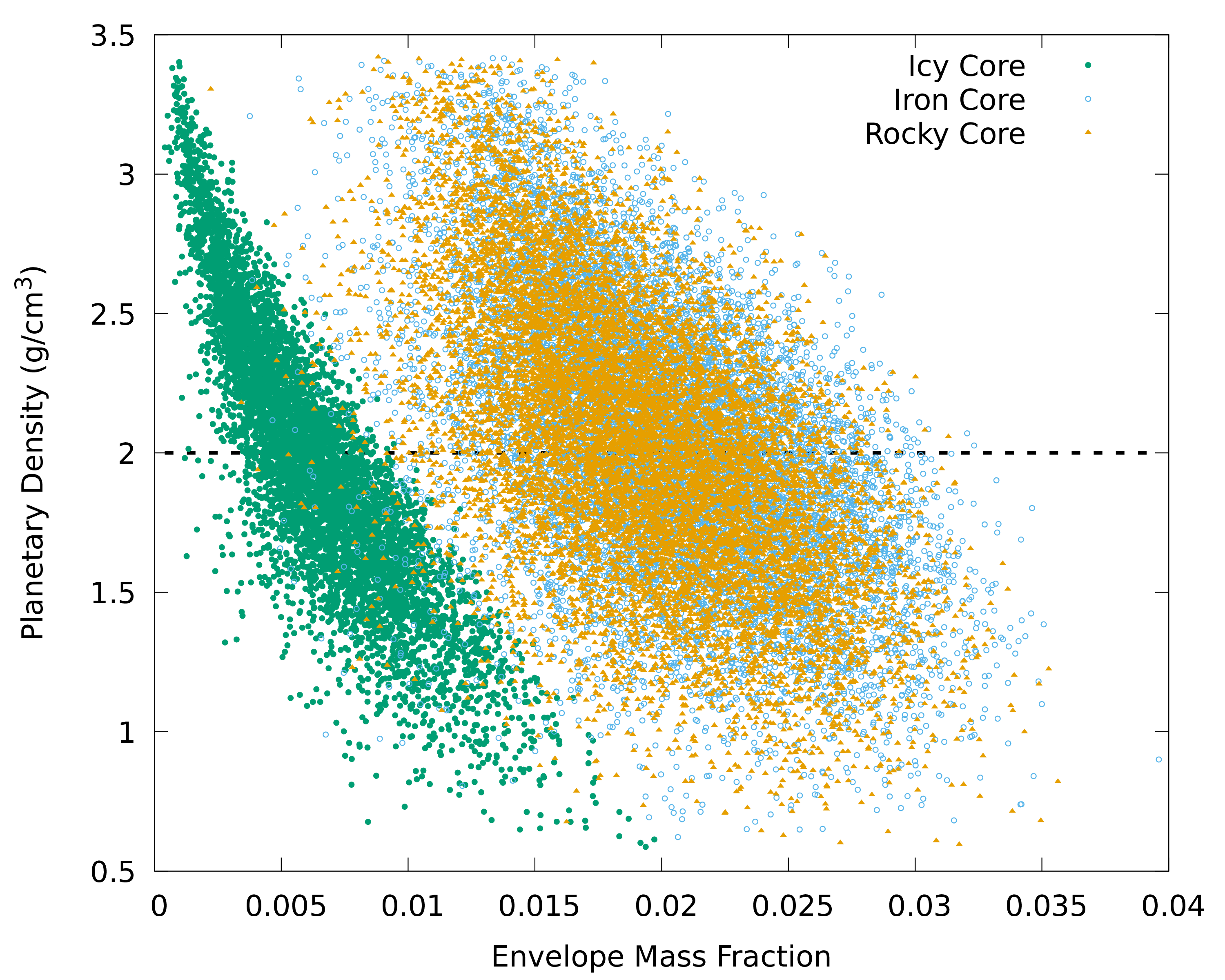}
    \caption{Monte Carlo investigation of the spread in the planetary density and the envelope mass fraction of planet~b resulting from the uncertainties on the stellar age, the planetary mass and radius, and the composition of the core under the assumption of enhanced metallicity of the planetary envelope. The three combinations of symbols and colors identify the three compositions and mass-radius relationships explored for the core, while the horizontal dashed line marks the retrieved bulk density of planet~b (without uncertainties for better clarity).}
    \label{fig:interiors_mc}
\end{figure}

The results of our Monte Carlo exploration are shown in Fig.\,\ref{fig:interiors_mc} for the case of enhanced envelope metallicity, which is likely more realistic for sub-Neptunian planets based on the case of the ice giants in the Solar System \citep[e.g.][and references therein]{DePAter2023} and the mass-metallicity relationships of giant planets from \citet{Thorngren2016}. The solutions shown are those that fall within 3$\sigma$ from the estimated planetary mass, radius, and density from Table \ref{tab:pparameters}. The results for the solar metallicity envelope are qualitatively identical. While the interior models adopted are less detailed than those discussed in Sect. \ref{sec:composition}, Fig.\,\ref{fig:interiors_mc} immediately shows that the characteristics of BD+00\,444\,b do not allow for core masses as small as those set by the pebble isolation mass shown in Fig. \ref{fig:tracks}.

As discussed by \citet{Mantovan2024} and \citet{Damasso2024}, however, the core masses set by the pebble isolation mass are actually lower limits because the population synthesis simulations do not account for two processes that can increase the final core mass. The first one is the accretion of planetesimals if the native protoplanetary disk was characterized by comparable abundances of pebbles and planetesimals, as it would allow the core to grow beyond the pebble isolation mass. The second one is the accretion of high-metallicity gas enriched by the volatiles released by the ices sublimating from the inward drifting pebbles if the protoplanetary disk was pebble-dominated. The high-metallicity gas could mimic the effects of a more massive core like those suggested by the interior studies of Jupiter \citep{Wahl2017,Stevenson2020} and Saturn \citep{Mankovich2021}.

The marked eccentricity of planet~b allows for a third possibility. While the candidate nature of planet~c and the lack of more detailed constraints on its orbit and mass hinder the quantification of the dynamical excitation of the systems \citep{Chambers2001,Turrini2020}, the orbital eccentricity of planet~b is highly suggestive of chaotic phases of dynamical evolution in its past \citep{Stuart1996,Chambers2001,Zinzi2017}. Unless its mass proves significantly higher than its currently estimated lower bound, the possible circular orbit for planet~c may indicate that it was not the planetary body that chaotically interacted with planet~b. 

Since collisions are the most likely outcome of chaotic evolution \citep{Chambers2001,Zinzi2017,Turrini2020}, the currently observed planet~b could be the result of the giant impact between two separate planetary bodies that reached the pebble isolation mass in a higher-multiplicity primordial planetary system. The higher core mass than what would be suggested by the pebble isolation mass profile alone could then be the result of the merging of the cores of these two planets. Collisions lead to the damping of the orbital eccentricity of the surviving body \citep{Chambers2001}, implying that the pre-collision eccentricity of planet b was higher than its current value. This, in turn, would have increased the likelihood of planetary collisions given the compact architecture of the system.

While these alternative scenarios for the formation environments and physical characteristics of BD+00\,444's planets cannot be discriminated based on the current data, future observations aimed at compositionally characterizing the atmosphere of planet~b have the potential to provide critical insight into the past of this system. If the two planets formed in a pebble-dominated disk, their atmospheres are expected to be highly enriched in the volatile elements C and O (and possibly N) and limited or not enriched at all in more refractory ones \citep{Booth2019,Turrini2021,Schneider2021}. 

If the two planets formed in planetesimal-rich disks or if planet~b is the result of a giant impact between two previously existing planets, the atmospheric composition of planet~b should show higher enrichment in S and refractories than in C and O \citep{Turrini2021,Pacetti2022}. Observational constraints on the S-over-N \citep{Turrini2021,Pacetti2022} or C-over-S \citep{Crossfield2023} abundance ratios could allow to shed light on the characteristics of BD+00\,444's formation history. Similarly, the scenarios dominated by mm-sized and cm-sized pebbles could be discriminated by constraining the enrichment of planet b in the volatile elements C, O, and N. Planets forming in disks dominated by cm-sized pebbles are expected to be characterized by enrichment in volatile elements of an order of magnitude or higher with respect to the host star, while mm-sized pebbles are expected to produce less marked enrichment \citep{Booth2019,Schneider2021}.

\begin{table}
    \caption{The input parameters used to run the MC modified {\tt GroMiT} code to produce simulated planets and investigate the formation history of TOI-2443\,$b$.}
    \label{tab:popsythesis}
    \centering
    \begin{tabular}{l c c}
        \hline \hline
    \multicolumn{2}{c}{Simulation parameters} \rule{0pt}{2.3ex} \rule[-1ex]{0pt}{0pt}\\
    \hline
      \multicolumn{1}{l}{N$^\circ$ of Monte Carlo runs} & \multicolumn{1}{c}{2$\times$10$^5$} \rule{0pt}{2.3ex} \rule[-1ex]{0pt}{0pt}\\
      \multicolumn{1}{l}{Seed formation time} & \multicolumn{1}{c}{0.0--3.0\,Myr} \\
      \multicolumn{1}{l}{Disk lifetime} & \multicolumn{1}{c}{$5.0$\,Myr} \\
    \hline
    \multicolumn{2}{c}{Star, planet \& disk properties} \rule{0pt}{2.3ex} \rule[-1ex]{0pt}{0pt}\\
    \hline
      \multicolumn{1}{l}{Stellar mass} & \multicolumn{1}{c}{0.642$\,$M${_\odot}$} \rule{0pt}{2.3ex} \rule[-1ex]{0pt}{0pt} \\
      \multicolumn{1}{l}{Stellar luminosity at 1\,Myr} & \multicolumn{1}{c}{0.85$\,$L${_\odot}$} \\
      \multicolumn{1}{l}{Disk mass} & \multicolumn{1}{c}{0.03$\,$M${_\odot}$} \\
      \multicolumn{1}{l}{Disk characteristic radius R$_c$} & \multicolumn{1}{c}{30.0$\,$au} \\
      \multicolumn{1}{l}{Surface density at R$_c$} & \multicolumn{1}{c}{6.05\,g cm$^{-2}$} \\
      \multicolumn{1}{l}{Temperature T$_0$ at R$_c$ } & \multicolumn{1}{c}{45$\,$K} \\
      \multicolumn{1}{l}{Disk accretion coefficient, $\alpha$} & \multicolumn{1}{c}{0.0003--0.003}\\
      \multicolumn{1}{l}{Turbulent viscosity, $\alpha$$_\nu$}& \multicolumn{1}{c}{0.0001}\\
      \multicolumn{1}{l}{Pebble size} & \multicolumn{1}{c}{1$\,$mm--1$\,$cm}\\
      \multicolumn{1}{l}{Dust-to-gas ratio} & \multicolumn{1}{c}{0.0075}\\
      \multicolumn{1}{l}{Seed mass} & \multicolumn{1}{c}{0.01$\,$M${_\oplus}$} \\
      \multicolumn{1}{l}{Initial envelope mass} & \multicolumn{1}{c}{0.0$\,$M${_\oplus}$} \\
      \multicolumn{1}{l}{Initial semimajor axis} & \multicolumn{1}{c}{0.1--30.0\,au} \\
    \hline
  \end{tabular}
  \label{tab:simparam}
\end{table}


\subsection{Tidal timescales and dissipation inside BD+00\,444\,b}
\label{tidal_analysis}
The large spatial ratio $a/R_{*}$ and small mass ratio $M_{\rm p}/M_{*}$ of planet~b make its tidal influence on the star negligible. In other words, it does not affect the orbit semi-major axis, stellar rotation, or the orientation of the stellar spin in any significant way. On the other hand, the tides raised by the star on planet~b are relevant for its rotation and internal dissipation of tidal energy. We adopted the constant time-lag model by \citet{Leconteetal10} where we express the product of the Love number $k_{2 \rm p}$ of the planet~by its tidal time lag $\Delta t_{\rm p}$ using the formula $k_{2 \rm p}\Delta t_{\rm p} = (2/3)Q^{\prime}_{\rm p}/n$, where $Q^{\prime}_{\rm p}$ is the modified tidal quality factor of the planet and $n = 2\pi/P_{\rm orb}$ is its orbital mean motion with $P_{\rm orb}$ being its orbital period. The advantage of a constant time-lag model is its validity for large values of eccentricity as in the case of BD+00\,444\,b.
 
The modified tidal quality factors of mini-Neptune planets are largely unknown. Using Uranus or Neptune as analogs for their tidal dissipation, one can adopt $Q^{\prime}_{\rm p}$ values between $4 \times 10^{4}$ and $1.5 \times 10^{5}$ and consider $k_{2\rm p}=$ 0.3--0.4 \citep{TittemoreWisdom90,ZhangHamilton08,Ogilvie14,JamesStixrude24}. Assuming for simplicity a value of $Q^{\prime}_{\rm p}=10^{5}$, we obtained an $e$-folding decay timescale of the orbital eccentricity $\tau_{\rm e_b} = 614$~Gyr for BD+00\,444\,b, much longer than Hubble time. This implies that tidal effects likely do not affect the eccentricities of planets b and c. The power dissipated by the tides inside planet~b is $P_{\rm tide} \sim 8 \times 10^{14}$~W for the present value of its eccentricity of $e_{\rm b}\approx$ 0.3. Both $P_{\rm tide}$ and $\tau_{\rm e_b}$ scale proportionally to $(Q^{\prime}_{\rm p})^{-1}$, so our estimate of $P_{\rm tide}$ is rather conservative. It is comparable to or even larger than the average extreme ultraviolet (XUV\footnote{XUV is the joint interval of extreme ultraviolet and soft X-rays (e.g. \citealt{Forcada2011}).}) flux received by the planet from its host star, which we estimated to be $L_{\rm XUV} \sim 3.2 \times 10^{14}$~W in the pass-band from 0.1 to 92 nm where photons can induce ionization and photo-evaporation of a hydrogen-rich planetary atmosphere. This estimate of $L_{\rm XUV}$ is based on the tables by \citet{Johnstoneetal21} considering a rotation period of the star of $\sim 45$~days, and is uncertain by at least a factor of a few due to the variable activity level at a given rotation period and the large uncertainty on the stellar rotation period (cf. Sect.~\ref{sec:periodogram}).

Another consequence of the eccentric orbit of planet~b is the pseudo-synchronization of its rotation, which is faster than the orbital period. For the present value of its eccentricity, the pseudo-synchronous rotation period is 10.1~days and the timescale for attaining pseudo-synchronization is of $\sim 2.5$~Myr for $Q^{\prime}_{\rm p}=10^{5}$ (see below for the indirect effect of the other planet).

The analysis presented above assumes the present value of the eccentricity of planet~b. In fact, the perturbations due to planet~c produce a secular exchange of angular momentum between the orbits, while their energies and their semi-major axes are secularly constant because this is a non-resonant system. The angular momentum exchanges induce a modulation of the eccentricities values. Unfortunately, the inclination of the planet's c orbit is unknown and its eccentricity only has an upper limit (Table\,\ref{tab:pparameters}). Therefore, a precise prediction of the amplitude and the period of the eccentricity modulations is not possible. A simplified treatment that can provide order-of-magnitude estimates is based on the model by \citet{Mardling07} that assumes co-planarity of the orbits of the two planets and a small eccentricity of the inner planet to allow the development of the secular perturbation equations to the first order in its eccentricity.

\begin{figure}
\centering
\includegraphics[width=0.5\textwidth]{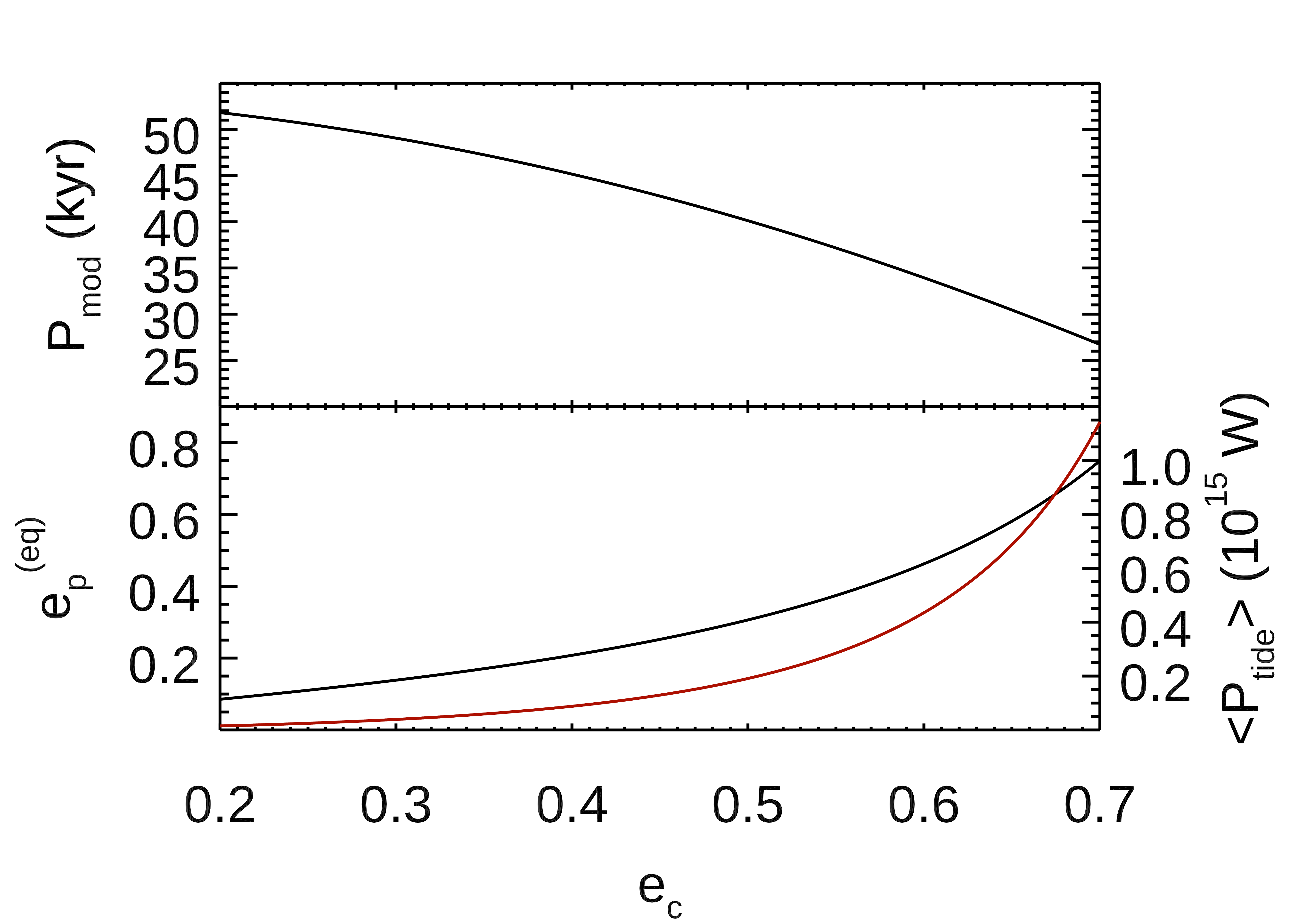}
\caption{Top panel: Circulation or libration period $P_{\rm mod}$ of the angle $\eta$ vs. the eccentricity $e_{\rm c}$ of the outer planet~c in the BD+00\,444 system. The  modulation of the eccentricity $e_{\rm b}$ of the orbit of the inner planet occurs with the same period. Lower panel: The equilibrium eccentricity $e_{\rm p}^{\rm (eq)}$ of the orbit of the inner planet~BD+00\,444\,b (black line, left axis) and the average tidal power $\langle P_{\rm tide} \rangle$ dissipated in its interior (red line, right axis) vs. the eccentricity of the orbit of the outer planet. }
    \label{eccentricity_plots}
\end{figure}

Adopting Mardling's simplifying hypotheses and considering that tides cannot significantly affect the eccentricities of the planetary orbits in our system, the angle $\eta = \varpi_{b} - \varpi_{c}$ between the arguments of periapsis of the two planets can be librating or circulating \citep[see Sect.~3 of][for details]{Mardling07}. The period of libration or circulation of $\eta$ is equal to the period of the modulation of the orbital eccentricity of planet~b, and is given in the top panel of Fig.~\ref{eccentricity_plots} as a function of the unknown value of the eccentricity of the outer planet~c. In any case, it turns out to be shorter than $\sim 6 \times 10^{4}$~yr, that is, much shorter than the pseudo-synchronization timescale of planet rotation. Therefore, we expect that the rotation of the planet synchronizes with the value corresponding to its average eccentricity. 

The short modulation period of the eccentricity $e_{\rm b}$ of planet~b implies that its present value may not be representative of the actual mean tidal dissipation inside the planet. In other words, being interested in the average value of the tidal power $P_{\rm diss}$ dissipated inside planet~b on evolutionary timescales, the fundamental parameter that we need is the mean value of its squared eccentricity $\langle e_{\rm b}^{2}\rangle$. The minimum value of $\langle e_{\rm b}^{2}\rangle $ is obtained when the angle $\eta$ circulates and is equal to $[e^{\rm (eq)}_{\rm p}]^2/2$, where $e_{\rm p}^{\rm (eq)}$ is the equilibrium eccentricity of planet~b as given by Eq.~(36) of \citet{Mardling07} and is a function of, among others, the unknown eccentricity of planet~c (see lower panel of Fig.~\ref{eccentricity_plots}, black line). The average dissipated power inside planet~b as a function of its equilibrium eccentricity is plotted in the lower panel of Fig.~\ref{eccentricity_plots} (red line) and is computed by means of Eqs.~(1) and (2) of \citet{Mardling07} adopting $Q_{\rm p}/k_{2\rm p} = 2Q^{\prime}_{\rm p}/3$ and $Q^{\prime}_{\rm p} = 10^{5}$. Mardling's second-order formula in the eccentricity provides systematically lower values of $P_{\rm diss}$ than the higher-order model by \citet{Leconteetal10}, which is in line with our estimate of a lower value for the tidal power and the poorly known value of $Q^{\prime}_{\rm p}$ for our planet.

Looking at the lower panel of Fig.~\ref{eccentricity_plots}, we conclude that the power dissipated by the tides inside planet~b becomes comparable with the XUV flux received from the star when its equilibrium eccentricity is of $\sim 0.3$ or larger. In turn, such an eccentricity of the inner planet~b implies a rather large eccentricity of the outer planet~c that, in our simplified model, must be larger than $\sim 0.5$, which is close to or above the lower limit as derived by the 2-planet model (see Sect.~\ref{sec:joint_analysis} and Table~\ref{tab:pparameters}).

Finally, we considered the case of a strong tidal dissipation inside planet~b, appropriate for a planet with a mainly solid interior, a possibility that cannot be excluded given the kind of structure models considered in Sect.~\ref{sec:composition}. Adopting $Q^{\prime}_{\rm p}=300$ \citep{Henningetal09}, the eccentricity $e$-folding decay time $\tau_{\rm e} \sim 1.9$~Gyr, while the tidal power dissipated inside the planet~becomes of $2.6 \times 10^{17}$~W for $e_{\rm b}=0.3$ corresponding to a surface heat flux of $\sim 90$~W~m$^{-2}$, that is, about 30 times larger than in the case of the Jupiter moon Io, suggesting a very strong volcanic activity potentially detectable by means of dedicated spectroscopic transit observations. In the presence of the outer planet~c, the strong tidal dissipation inside planet~b would drive its eccentricity toward its equilibrium value $e^{\rm (eq)}_{\rm p}$ over a timescale of a few $\tau_{\rm e}$'s, comparable with the age of the system, given that BD+00\,444 is likely to be an old star \citep[see Sect.~3.3 of][]{Mardling07}. In this scenario, the current value of the eccentricity would be close to the equilibrium value implying a rather large eccentricity of planet~c ($e_{\rm c} \sim 0.5$, cf. Fig.~\ref{eccentricity_plots}, lower panel) and an extended phase of remarkable tidal dissipation inside planet~b with an average $P_{\rm diss}$ of the order of $10^{17}$~W. Such predictions of a huge internal tidal heating in planet~b and of a rather large eccentricity of planet~c can provide observational tests for such a high tidal dissipation regime resulting from a low $Q^{\prime}_{\rm p}$ value.

\label{sec:TSM}
\subsection{Atmospheric prospects for BD+00\,444\,b}

\begin{figure}
    \centering
    \includegraphics[width=0.5\textwidth]{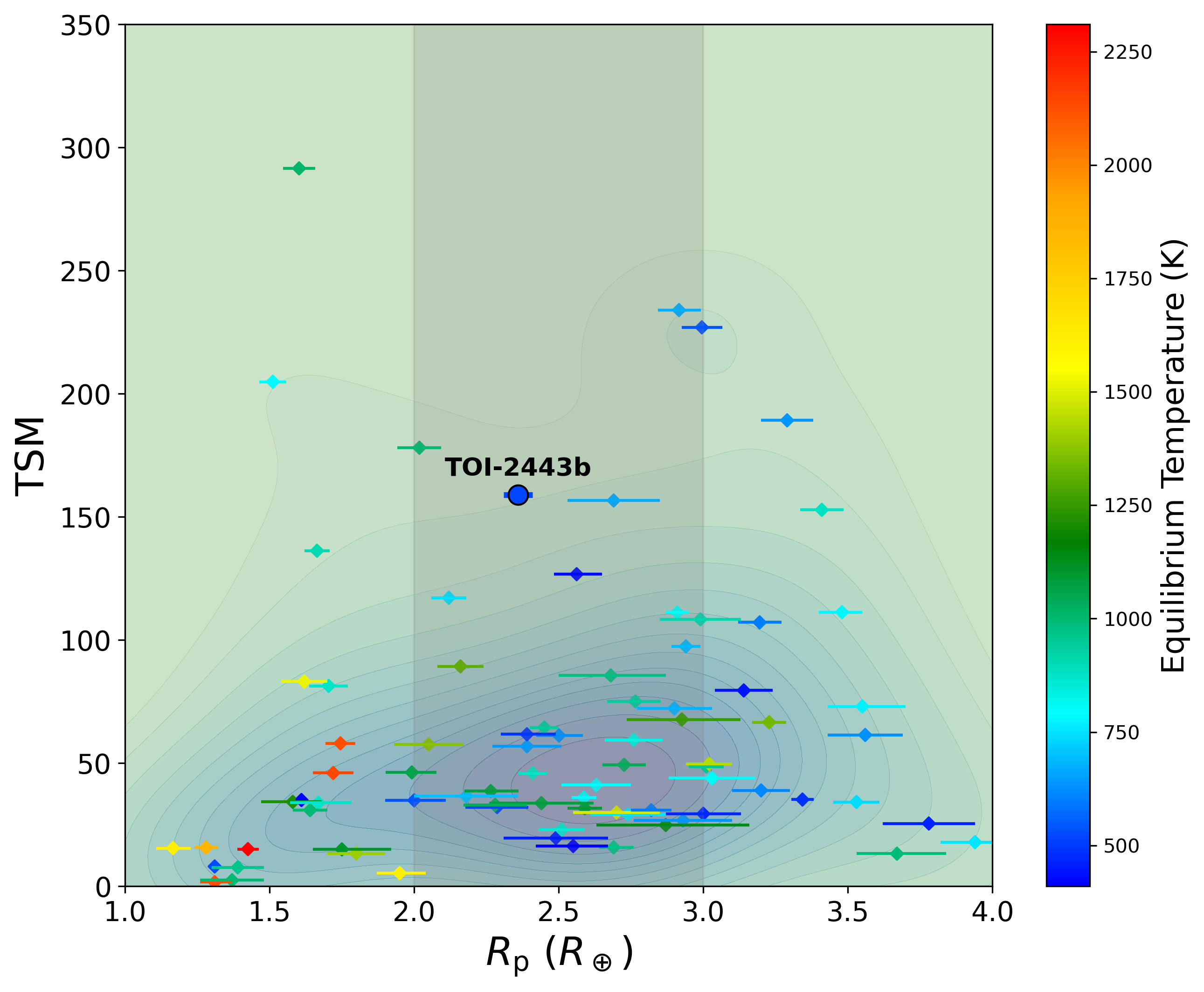}
    \caption{TSM values over the radius of planets with well-characterized radius and mass (with 5$\sigma$, 4$\sigma$ accuracy, respectively) orbiting around FGK stars. The color of the points represents their equilibrium temperature, while the TSM error bars are suppressed for clarity.}
    \label{fig:TSM}
\end{figure}

BD+00\,444\,b's exceptionally high TSM of $159^{+46}_{-31}$ (Table\,\ref{tab:pparameters}) makes it an ideal candidate for atmospheric characterization with JWST. This metric indicates that its atmosphere is well-suited for in-depth study via transmission spectroscopy, offering an opportunity to detect key molecular features. Interestingly, only three other sub-Neptunes with $2\,R_{\oplus}<R_p<3\,R_{\oplus}$, orbiting around FGK stars, have higher TSM values (i.e. HD\,136352\,c, HD\,191939\,d, and TOI-544\,b, respectively from \citealt{Delrez2021}, \citealt{Orell-Miquel2023} and \citealt{Osborne2024}), as depicted in Fig.\,\ref{fig:TSM}, to the best of our knowledge.

The highly dissipative tidal scenario (cf. Sect.~\ref{sec:formation}) suggests strong volcanic activity on planet~b, which could potentially be detected through spectroscopic observations during transits. In this scenario, the eccentricity of BD+00\,444\,b would be maintained by perturbations by candidate planet~c, whose orbit would also be highly eccentric ($e_{\rm c} \sim 0.5$). Alternatively, the rather large orbital eccentricity of planet~b might have also originated from a primordial impact and persisted due to rheological properties similar to that of Uranus or Neptune, resulting in minimal eccentricity damping over time (cf. Sect.~\ref{tidal_analysis}). Furthermore, investigating the atmosphere of BD+00\,444\,b would allow us to constrain its water content, which in our models can span between null and 66\%, and learn whether the planet is rich in volatile elements (likely a consequence of a cm-sized pebble-dominated formation disk), while comparing S, C, and O enrichment may provide a clue on whether BD+00\,444\,b formed in a planetesimal-rich disk and possibly underwent a giant impact.

\section{Summary}
\label{sec:conclusions}
%
%
Thanks to precise RV measurements obtained with the high-resolution spectrograph HARPS-N, we determined the mass of BD+00\,444\,b to be $4.8 \pm 1.1 \, M_{\oplus}$. With a radius of $2.36 \pm 0.05 \, R_{\oplus}$, this transiting sub-Neptune, initially discovered by \emph{TESS}, is in an eccentric orbit, $e=0.301_{-0.034}^{+0.046}$, of about 15.67 days around a quiet K5\,V star (Table\,\ref{tab:star}), and has a bulk density of $2.00^{+0.49}_{-0.45}$ g cm$^{-3}$. In particular, BD+00\,444\,b turns out to have a high TSM of $159^{+46}_{-31}$ and, therefore, it is an optimal candidate for atmospheric follow-up with JWST.

While we can exclude the presence of Neptune-mass planets up to 1 au, and Jupiter-mass planets up to 10 au (Fig.\,\ref{fig:completeness}), unless severely inclined with respect to the orbital plane of BD+00\,444\,b, our analysis does suggest the existence of another planet. Candidate BD+00\,444\,c, on a longer orbital period of about 97 days ($\sim$ 0.4 au), has an equilibrium temperature of $283\pm4$ K that would place it within the habitable zone, and a minimum mass, $M_{\rm c}\sin{i}$, of $9.3^{+1.8}_{-2.0}\,M_{\oplus}$, which likely lies within a factor of 2 of the true one (i.e. it is hardly more massive than Neptune). However, we expect this candidate to be non-transiting and therefore it will necessitate more RV measurements to be secured with higher statistical evidence.  

\begin{acknowledgements}
We would like to thank the referee J.\,A. Caballero for giving many constructive comments that substantially helped improving the quality of the paper. We acknowledge financial contribution from the INAF Large Grant 2023 ``EXODEMO''. This work is based on observations made with the Italian Telescopio Nazionale Galileo (TNG) operated by the Fundaci\'{o}n Galileo Galilei (FGG) of the Istituto Nazionale di Astrofisica (INAF) at the Observatorio del Roque de los Muchachos (La Palma, Canary Islands, Spain). We acknowledge the Italian center for Astronomical Archives (IA2, \url{https://www.ia2.inaf.it}), part of the Italian National Institute for Astrophysics (INAF), for providing technical assistance, services and supporting activities of the GAPS collaboration. This work includes data collected with the TESS mission, obtained from the MAST data archive at the Space Telescope Science Institute (STScI). Funding for the TESS mission is provided by the NASA Explorer Program. STScI is operated by the Association of Universities for Research in Astronomy, Inc., under NASA contract NAS 5–26555. We acknowledge the use of public TESS data from pipelines at the TESS Science Office and at the TESS Science Processing Operations Center. Resources supporting this work were provided by the NASA High-End Computing (HEC) Program through the NASA Advanced Supercomputing (NAS) Division at Ames Research Center for the production of the SPOC data products. Funding for the TESS mission is provided by NASA's Science Mission Directorate. KAC and CNW acknowledge support from the TESS mission via sub-award s3449 from MIT.
This work makes use of observations from the LCOGT network. Part of the LCOGT telescope time was granted by NOIRLab through the Mid-Scale Innovations Program (MSIP). MSIP is funded by NSF.
This research made use of the Exoplanet Follow-up Observation Program (ExoFOP; DOI: 10.26134/ExoFOP5) website, which is operated by the California Institute of Technology, under contract with the National Aeronautics and Space Administration under the Exoplanet Exploration Program.
The research made use of the SIMBAD database, operated at CDS, Strasbourg, France, NASA’s Astrophysics Data System and the NASA Exoplanet Archive, which is operated by the California Institute of Technology, under contract with the National Aeronautics and Space Administration under the Exoplanet Exploration Program.
The work is based on observations made with the Italian Telescopio Nazionale Galileo (TNG) operated on the island of La Palma by the Fundacion Galileo Galilei of the INAF (Istituto Nazionale di Astrofisica) at the Spanish Observatorio del Roque de los Muchachos of the Instituto de Astrofisica de Canarias. 
This publication makes use of The Data \& Analysis Center for Exoplanets (DACE), which is a facility based at the University of Geneva (CH) dedicated to extrasolar planets data visualization, exchange and analysis. DACE is a platform of the Swiss National Centre of Competence in Research (NCCR) PlanetS, federating the Swiss expertise in Exoplanet research. The DACE platform is available at \url{https://dace.unige.ch}. This work also made use of data from the European Space Agency (ESA) mission Gaia (\url{https://www.cosmos.esa.int/gaia}), processed by the Gaia Data Processing and Analysis Consortium (DPAC, \url{https://www.cosmos.esa.int/web/gaia/dpac/consortium}). 
L.M. acknowledges financial contribution from PRIN MUR 2022 project 2022J4H55R. D.P. acknowledges the support from the Istituto Nazionale di Oceanografia e Geofisica Sperimentale (OGS) and CINECA through the program ``HPC-TRES (High Performance Computing Training and Research for Earth Sciences)'' award number 2022-05 as well as the support of the PRIN INAF 2019 ``Planetary systems at young ages (PLATEA)''. D.T. and D.P. acknowledge the support from the ASI-INAF grant no. 2021-5-HH.0 plus addenda no. 2021-5-HH.1-2022 and 2021-5-HH.2-2024 as well as the computational support of the Genesis cluster at INAF-IAPS. D.T. acknowledges the support of the European Research Council via the Horizon 2020 Framework Programme ERC Synergy ``ECOGAL'' Project GA-855130.
M.P. acknowledges support from the European Union – NextGenerationEU (PRIN MUR 2022 20229R43BH) and the ``Programma di Ricerca Fondamentale INAF 2023''.
C.D. and M.S. acknowledge support from the Swiss National Science Foundation under grant TMSGI2\_211313.
X.D. acknowledges the support from the European Research Council (ERC) under the European Union's Horizon 2020 research and innovation programme (grant agreement SCORE No 851555) and from the Swiss National Science Foundation under the grant SPECTRE (No 200021\_215200).
\end{acknowledgements}

%
%

\bibliographystyle{aa}
\bibliography{2443}

%
%

\appendix
\section{HARPS-N RV datapoints}\label{appendix:rv}
\begin{table*}
\centering
\caption[]{HARPS-N RV data points and activity indices obtained with the DRS pipeline.}\label{tab:rv}
{\tiny\renewcommand{\arraystretch}{.8}
\resizebox{!}{.4\paperheight}{%
\begin{tabular}{llllllllllll}
    \hline\hline
    $\mathrm{BJD_{\textsf{TDB}}}$ & RV & $\pm1\upsigma_{\textsf{RV}}$ & FWHM & $\upsigma_{\textsf{FWHM}}$ & BIS$^{(\dagger)}$ & $\upsigma_{\textsf{BIS}}$ & Contrast & $\upsigma_{\textsf{Cont}}$ & $S_{MW}$ & $\upsigma_{S_{MW}}$ \\ 
    $-2457000\,[d]$ & $\mathrm{~~[m\,s^{-1}]}$ & $\mathrm{~~[m\,s^{-1}]}$ & & & & & & & & & \rule[-0.8ex]{0pt}{0pt} \\ 
    \hline \\
2413.723798 & 72988.38 & 1.51 & 5497.89 & 3.43 & 29.81 & 3.43 & 54.452 & 0.034 & 0.4220 & 0.0034 \\
2414.726264 & 72991.49 & 1.00 & 5502.51 & 2.16 & 30.64 & 2.16 & 54.529 & 0.021 & 0.4224 & 0.0017 \\
2416.694371 & 72989.73 & 1.52 & 5493.69 & 3.47 & 26.09 & 3.47 & 54.503 & 0.034 & 0.4271 & 0.0037 \\
2418.698307 & 72985.10 & 2.38 & 5511.22 & 5.80 & 32.13 & 5.80 & 54.350 & 0.057 & 0.4087 & 0.0073 \\
2429.737786 & 72982.75 & 2.04 & 5492.84 & 4.89 & 24.94 & 4.89 & 54.481 & 0.048 & 0.3479 & 0.0052 \\
2443.74442 & 72983.92 & 1.34 & 5493.68 & 3.00 & 30.43 & 3.00 & 54.552 & 0.030 & 0.3997 & 0.0027 \\
2444.715427 & 72986.25 & 1.65 & 5491.94 & 3.78 & 37.59 & 3.78 & 54.576 & 0.038 & 0.3893 & 0.0039 \\
2445.725559 & 72986.73 & 0.81 & 5491.02 & 1.72 & 28.65 & 1.72 & 54.624 & 0.017 & 0.4108 & 0.0011 \\
2446.734648 & 72985.33 & 0.82 & 5498.10 & 1.73 & 29.80 & 1.73 & 54.578 & 0.017 & 0.4095 & 0.0011 \\
2447.718026 & 72982.38 & 0.78 & 5495.09 & 1.64 & 27.73 & 1.64 & 54.582 & 0.016 & 0.4176 & 0.0010 \\
2448.724215 & 72984.23 & 0.64 & 5492.88 & 1.32 & 31.00 & 1.32 & 54.584 & 0.013 & 0.4215 & 0.0007 \\
2449.720491 & 72981.30 & 0.98 & 5493.92 & 2.12 & 26.94 & 2.12 & 54.593 & 0.021 & 0.4042 & 0.0016 \\
2456.753204 & 72987.60 & 0.62 & 5494.33 & 1.30 & 30.15 & 1.30 & 54.583 & 0.013 & 0.4183 & 0.0007 \\
2457.639221 & 72984.14 & 0.71 & 5493.84 & 1.49 & 28.70 & 1.49 & 54.588 & 0.015 & 0.4289 & 0.0009 \\
2458.752784 & 72985.28 & 0.89 & 5499.23 & 1.91 & 26.58 & 1.91 & 54.559 & 0.019 & 0.4194 & 0.0013 \\
2459.665599 & 72985.35 & 0.96 & 5495.67 & 2.05 & 29.85 & 2.05 & 54.586 & 0.020 & 0.3988 & 0.0015 \\
2460.741176 & 72981.96 & 0.76 & 5492.35 & 1.61 & 28.54 & 1.61 & 54.582 & 0.016 & 0.4258 & 0.0010 \\
2461.69778 & 72983.97 & 0.67 & 5494.82 & 1.40 & 30.56 & 1.40 & 54.590 & 0.014 & 0.4150 & 0.0008 \\
2462.658313 & 72984.96 & 1.01 & 5497.65 & 2.20 & 34.16 & 2.20 & 54.536 & 0.022 & 0.4125 & 0.0016 \\
2463.726227 & 72981.74 & 1.04 & 5493.15 & 2.25 & 29.00 & 2.25 & 54.565 & 0.022 & 0.4063 & 0.0018 \\
2464.644554 & 72987.49 & 1.79 & 5497.30 & 4.17 & 32.10 & 4.17 & 54.546 & 0.041 & 0.4075 & 0.0045 \\
2465.656687 & 72982.52 & 0.84 & 5497.88 & 1.77 & 27.73 & 1.77 & 54.583 & 0.018 & 0.4127 & 0.0012 \\
2472.717053 & 72989.24 & 0.94 & 5497.79 & 2.02 & 28.16 & 2.02 & 54.542 & 0.020 & 0.4112 & 0.0015 \\
2473.705081 & 72990.06 & 1.98 & 5503.97 & 4.68 & 30.58 & 4.68 & 54.505 & 0.046 & 0.3755 & 0.0052 \\
2475.694546 & 72985.17 & 1.94 & 5489.70 & 4.61 & 30.22 & 4.61 & 54.519 & 0.046 & 0.4167 & 0.0052 \\
2476.698698 & 72985.64 & 0.72 & 5496.54 & 1.51 & 29.78 & 1.51 & 54.591 & 0.015 & 0.4148 & 0.0009 \\
2477.682098 & 72987.47 & 0.82 & 5493.95 & 1.73 & 27.50 & 1.73 & 54.578 & 0.017 & 0.4048 & 0.0011 \\
2478.659949 & 72985.71 & 0.90 & 5499.79 & 1.93 & 34.02 & 1.93 & 54.588 & 0.019 & 0.4202 & 0.0014 \\
2479.633364 & 72987.95 & 0.73 & 5496.68 & 1.53 & 27.08 & 1.53 & 54.572 & 0.015 & 0.4223 & 0.0009 \\
2481.546474 & 72983.02 & 1.38 & 5502.62 & 3.13 & 27.87 & 3.13 & 54.575 & 0.031 & 0.4006 & 0.0030 \\
2516.622614 & 72986.91 & 0.95 & 5467.09 & 2.04 & 30.70 & 2.04 & 54.835 & 0.021 & 0.3952 & 0.0018 \\
2579.485244 & 72988.66 & 0.96 & 5473.07 & 2.05 & 26.55 & 2.05 & 54.784 & 0.021 & 0.4131 & 0.0018 \\
2580.474733 & 72988.77 & 1.29 & 5469.46 & 2.86 & 30.54 & 2.86 & 54.765 & 0.029 & 0.3980 & 0.0029 \\
2584.465247 & 72990.79 & 4.37 & 5505.06 & 11.72 & 29.93 & 11.72 & 54.479 & 0.116 & 0.1831 & 0.0189 \\
2586.471444 & 72997.16 & 3.16 & 5491.58 & 8.05 & 33.43 & 8.05 & 54.486 & 0.080 & 0.3079 & 0.0124 \\
2599.369051 & 72988.06 & 6.55 & 5454.92 & 18.33 & 19.52 & 18.33 & 54.691 & 0.184 & 0.1912 & 0.0309 \\
2601.428013 & 72991.87 & 0.79 & 5465.03 & 1.68 & 27.13 & 1.68 & 54.932 & 0.017 & 0.3794 & 0.0012 \\
2615.38517 & 72987.88 & 0.93 & 5470.37 & 1.98 & 27.45 & 1.98 & 54.860 & 0.020 & 0.3857 & 0.0016 \\
2625.359895 & 72984.79 & 0.81 & 5473.12 & 1.71 & 24.34 & 1.71 & 54.820 & 0.017 & 0.4239 & 0.0013 \\
2626.369006 & 72986.51 & 1.24 & 5476.47 & 2.75 & 26.75 & 2.75 & 54.750 & 0.027 & 0.4077 & 0.0029 \\
2627.345312 & 72985.86 & 2.08 & 5471.31 & 4.94 & 24.86 & 4.94 & 54.680 & 0.049 & 0.4100 & 0.0066 \\
2628.338867 & 72993.33 & 2.00 & 5472.57 & 4.73 & 20.03 & 4.73 & 54.661 & 0.047 & 0.4175 & 0.0064 \\
2629.330844 & 72990.21 & 0.88 & 5479.55 & 1.88 & 28.70 & 1.88 & 54.721 & 0.019 & 0.4361 & 0.0015 \\
2638.336484 & 72980.80 & 1.13 & 5469.52 & 2.46 & 30.33 & 2.46 & 54.801 & 0.025 & 0.4243 & 0.0024 \\
2789.716288 & 72989.36 & 2.91 & 5468.83 & 7.28 & 34.94 & 7.28 & 54.706 & 0.073 & 0.2655 & 0.0109 \\
2792.718815 & 72985.54 & 1.34 & 5474.78 & 2.98 & 25.77 & 2.98 & 54.799 & 0.030 & 0.4009 & 0.0032 \\
2801.687049 & 72987.14 & 1.09 & 5470.20 & 2.36 & 30.68 & 2.36 & 54.799 & 0.024 & 0.4023 & 0.0021 \\
2802.683665 & 72986.87 & 0.72 & 5468.85 & 1.50 & 28.56 & 1.50 & 54.845 & 0.015 & 0.4064 & 0.0010 \\
2803.688393 & 72985.10 & 0.84 & 5471.45 & 1.78 & 29.15 & 1.78 & 54.828 & 0.018 & 0.4023 & 0.0013 \\
2804.66552 & 72985.73 & 0.98 & 5468.98 & 2.10 & 28.18 & 2.10 & 54.856 & 0.021 & 0.4010 & 0.0018 \\
2816.715226 & 72991.79 & 1.37 & 5476.37 & 3.09 & 27.16 & 3.09 & 54.756 & 0.031 & 0.4235 & 0.0032 \\
2821.682561 & 72987.44 & 1.43 & 5470.75 & 3.22 & 25.45 & 3.22 & 54.726 & 0.032 & 0.3938 & 0.0033 \\
2822.741327 & 72986.73 & 1.36 & 5483.40 & 3.04 & 29.35 & 3.04 & 54.698 & 0.030 & 0.4118 & 0.0031 \\
2830.661198 & 72983.43 & 1.13 & 5465.13 & 2.47 & 29.96 & 2.47 & 54.869 & 0.025 & 0.4111 & 0.0024 \\
2831.604728 & 72984.40 & 1.21 & 5475.31 & 2.67 & 23.73 & 2.67 & 54.782 & 0.027 & 0.4003 & 0.0027 \\
2832.610799 & 72985.38 & 1.16 & 5473.59 & 2.55 & 24.36 & 2.55 & 54.814 & 0.025 & 0.4293 & 0.0024 \\
2833.723387 & 72985.65 & 1.20 & 5477.27 & 2.65 & 25.34 & 2.65 & 54.776 & 0.026 & 0.4118 & 0.0024 \\
2834.568635 & 72987.69 & 1.10 & 5471.28 & 2.39 & 29.76 & 2.39 & 54.772 & 0.024 & 0.4123 & 0.0023 \\
2843.580977 & 72983.92 & 0.77 & 5482.36 & 1.62 & 28.71 & 1.62 & 54.684 & 0.016 & 0.4353 & 0.0011 \\
2844.588798 & 72984.64 & 0.81 & 5483.02 & 1.71 & 28.93 & 1.71 & 54.694 & 0.017 & 0.4400 & 0.0013 \\
2845.585391 & 72984.75 & 0.82 & 5482.81 & 1.74 & 29.46 & 1.74 & 54.699 & 0.017 & 0.4474 & 0.0013 \\
2858.544072 & 72983.70 & 1.58 & 5479.43 & 3.60 & 35.24 & 3.60 & 54.772 & 0.036 & 0.4176 & 0.0039 \\
2859.531703 & 72983.60 & 0.97 & 5471.31 & 2.07 & 30.39 & 2.07 & 54.812 & 0.021 & 0.3987 & 0.0018 \\
2860.525588 & 72982.16 & 0.92 & 5466.72 & 1.97 & 30.72 & 1.97 & 54.847 & 0.020 & 0.4072 & 0.0016 \\
2861.52357 & 72985.92 & 0.96 & 5464.30 & 2.05 & 30.61 & 2.05 & 54.862 & 0.021 & 0.4022 & 0.0017 \\
2864.53456 & 72985.99 & 1.12 & 5468.24 & 2.45 & 24.19 & 2.45 & 54.866 & 0.025 & 0.3955 & 0.0023 \\
2866.679822 & 72988.91 & 2.02 & 5477.48 & 4.77 & 21.57 & 4.77 & 54.779 & 0.048 & 0.4055 & 0.0060 \\
2867.659865 & 72988.69 & 1.03 & 5470.03 & 2.23 & 32.50 & 2.23 & 54.824 & 0.022 & 0.4070 & 0.0019 \\
2868.61734 & 72988.06 & 0.84 & 5465.68 & 1.78 & 27.55 & 1.78 & 54.853 & 0.018 & 0.3921 & 0.0013 \\
2869.630073 & 72986.27 & 0.88 & 5468.86 & 1.86 & 27.38 & 1.86 & 54.824 & 0.019 & 0.4046 & 0.0014 \\
2870.588191 & 72985.01 & 1.22 & 5470.95 & 2.68 & 27.62 & 2.68 & 54.822 & 0.027 & 0.3913 & 0.0026 \\
2871.496933 & 72984.56 & 1.10 & 5467.31 & 2.38 & 28.31 & 2.38 & 54.852 & 0.024 & 0.3879 & 0.0022 \\
2872.470285 & 72985.00 & 1.35 & 5465.48 & 3.00 & 32.78 & 3.00 & 54.875 & 0.030 & 0.4088 & 0.0033 \\
2873.582671 & 72982.97 & 0.85 & 5466.93 & 1.81 & 29.11 & 1.81 & 54.862 & 0.018 & 0.3977 & 0.0013 \\
2893.564691 & 72991.30 & 0.78 & 5484.95 & 1.64 & 24.03 & 1.64 & 54.625 & 0.016 & 0.4534 & 0.0011 \\
2894.591783 & 72992.30 & 1.17 & 5492.45 & 2.55 & 30.65 & 2.55 & 54.582 & 0.025 & 0.4566 & 0.0023 \\
2895.403516 & 72993.60 & 0.76 & 5488.14 & 1.59 & 30.76 & 1.59 & 54.585 & 0.016 & 0.4496 & 0.0011 \\
2899.437509 & 72989.29 & 1.43 & 5484.33 & 3.21 & 33.85 & 3.21 & 54.582 & 0.032 & 0.4281 & 0.0035 \\
2909.420773 & 72984.67 & 1.04 & 5476.22 & 2.25 & 29.10 & 2.25 & 54.759 & 0.023 & 0.3955 & 0.0020 \\
2911.547608 & 72983.97 & 1.02 & 5470.54 & 2.19 & 30.30 & 2.19 & 54.797 & 0.022 & 0.3963 & 0.0019 \\
2912.498629 & 72986.82 & 2.63 & 5476.73 & 6.48 & 25.63 & 6.48 & 54.586 & 0.065 & 0.4061 & 0.0093 \\
2913.572334 & 72984.72 & 0.96 & 5466.27 & 2.06 & 29.60 & 2.06 & 54.831 & 0.021 & 0.4139 & 0.0017 \\
2931.494634 & 72984.33 & 0.92 & 5469.56 & 1.96 & 28.18 & 1.96 & 54.781 & 0.020 & 0.4212 & 0.0016 \\
2932.486548 & 72982.42 & 1.40 & 5473.30 & 3.13 & 29.28 & 3.13 & 54.731 & 0.031 & 0.4252 & 0.0034 \\
2934.511932 & 72984.36 & 0.77 & 5482.92 & 1.62 & 27.57 & 1.62 & 54.665 & 0.016 & 0.4467 & 0.0012 \\
2936.526519 & 72985.29 & 0.82 & 5487.66 & 1.72 & 24.72 & 1.72 & 54.588 & 0.017 & 0.4528 & 0.0013 \\
2938.430658 & 72988.96 & 0.70 & 5491.45 & 1.46 & 27.32 & 1.46 & 54.569 & 0.015 & 0.4580 & 0.0009 \\
2950.441448 & 72982.43 & 0.98 & 5470.54 & 2.11 & 30.35 & 2.11 & 54.780 & 0.021 & 0.4168 & 0.0018 \\
2951.472841 & 72985.04 & 1.30 & 5472.64 & 2.88 & 31.30 & 2.88 & 54.775 & 0.029 & 0.4167 & 0.0031 \\
2952.41688 & 72983.46 & 1.15 & 5473.31 & 2.53 & 31.97 & 2.53 & 54.801 & 0.025 & 0.4202 & 0.0025 \\
2953.408719 & 72987.56 & 3.48 & 5468.93 & 8.94 & 29.29 & 8.94 & 54.650 & 0.089 & 0.2588 & 0.0140 \\
2954.379789 & 72985.40 & 5.00 & 5485.61 & 13.56 & 30.64 & 13.56 & 54.698 & 0.135 & 0.5340 & 0.0230 \\
2958.436163 & 72991.06 & 2.07 & 5463.16 & 4.92 & 25.87 & 4.92 & 54.808 & 0.049 & 0.3737 & 0.0066 \\
2959.454356 & 72989.14 & 1.29 & 5461.36 & 2.86 & 27.22 & 2.86 & 54.913 & 0.029 & 0.4044 & 0.0031 \\
2960.402022 & 72987.99 & 2.72 & 5452.66 & 6.75 & 32.61 & 6.75 & 54.899 & 0.068 & 0.3766 & 0.0100 \\
2962.399155 & 72984.87 & 1.32 & 5466.43 & 2.93 & 24.75 & 2.93 & 54.913 & 0.029 & 0.3746 & 0.0031 \\
2963.454805 & 72985.34 & 1.55 & 5462.62 & 3.51 & 31.77 & 3.51 & 54.893 & 0.035 & 0.3886 & 0.0042 \rule[-0.8ex]{0pt}{0pt} \\
    \bottomrule
\end{tabular}}}
\end{table*}

\begin{table*}
\centering
\caption{Priors and posteriors for the parameters of the preferred joint model fit of Sect.~\ref{sec:joint_analysis}.}\label{tab:prior2p}
\renewcommand{\arraystretch}{1.2}\begin{tabular}{lllc}
    \hline\hline
    \multicolumn{2}{l}{Parameter} &  Prior distribution & Posterior values \\ 
    \hline
    \multicolumn{2}{l}{Keplerian Parameters:} & \rule{0pt}{2.6ex} \\ 
    $\rho_{\star}$ & $\mathrm{[g\,cm^{-3}]}$   & $\mathcal{N}(3.61,\:0.24)$ & Table\,\ref{tab:star} \\
    $T_{0,b}$ & $\mathrm{[BJD_{TDB}]}$    & $\mathcal{N}(2459148.099,\:0.001)$ & Table\,\ref{tab:pparameters} \\
    $P_b$ & $\mathrm{[d]}$   & $\mathcal{N}(15.669,\:0.001)$ & Table\,\ref{tab:pparameters} \\
    $T_{0,c}$ & $\mathrm{[BJD_{TDB}]}$    & $\mathcal{U}(2459144,\:2459275)$ & Table\,\ref{tab:pparameters} \\
    $P_c$ & $\mathrm{[d]}$   & $\mathcal{U}(20,\:150)$ & Table\,\ref{tab:pparameters} \\
    $(e_b, e_c)^*$ &  & $0$ & Table\,\ref{tab:pparameters} \\
    $(\omega_b, \omega_c)^*$ &  & $90$ & Table\,\ref{tab:pparameters} \\
    
    \multicolumn{2}{l}{Transit Parameters:} & \rule{0pt}{2.6ex} \\ 
    $R_{\rm p}/R_{\rm \star}$ &     & $\mathcal{U}(0.0,\:1.0)$ & Table\,\ref{tab:pparameters} \\
    $D$ &    & $1.0$ & -- \\
    $q_1$ (TESS) &     & $\mathcal{T}(0.45,\:0.27)$ & Table\,\ref{tab:pparameters} \\
    $q_2$ (TESS) &     & $\mathcal{T}(0.38,\:0.30)$ & Table\,\ref{tab:pparameters} \\
    $q_1$ (LCO$_Y$) &     & $\mathcal{U}(0,\:1)$ & Table\,\ref{tab:pparameters} \\
    $q_2$ (LCO$_Y$) &     & $\mathcal{U}(0,\:1)$ & Table\,\ref{tab:pparameters} \\
    $q_1$ (LCO$_{zs}$) &     & $\mathcal{U}(0,\:1)$ & Table\,\ref{tab:pparameters} \\
    $q_2$ (LCO$_{zs}$) &     & $\mathcal{U}(0,\:1)$ & Table\,\ref{tab:pparameters} \\

    \multicolumn{3}{l}{Light curve GP Hyperparameters:} & \rule{0pt}{2.6ex} \\ 
    $\sigma_{\textsf{TESS}}$ & $\mathrm{[ppt]}$  & $\mathcal{L}(10^{-5},\:1)$ & Table\,\ref{tab:pparameters} \\
    $\rho_{\textsf{TESS}}$ & $\mathrm{[d]}$  & $\mathcal{L}(10^{-1},\:100)$ & Table\,\ref{tab:pparameters} \\
    
    \multicolumn{2}{l}{\footnotesize{RV parameters:}} \rule{0pt}{2.6ex} \\
    $K_b$ & $\mathrm{[m\,s^{-1}]}$   & $\mathcal{U}(0,\:10)$ & Table\,\ref{tab:pparameters} \\
    $K_c$ & $\mathrm{[m\,s^{-1}]}$   & $\mathcal{U}(0,\:10)$ & Table\,\ref{tab:pparameters} \\
    $\overline{\mu}_{\textsf{HARPS-N}}$ & $\mathrm{[m\,s^{-1}]}$ & $\mathcal{U}(72975,\:72995)$ & Table\,\ref{tab:pparameters}\\
    $\sigma_{\textsf{HARPS-N}}$ & $\mathrm{[m\,s^{-1}]}$   & $\mathcal{U}(0,\:10)$ & Table\,\ref{tab:pparameters} \\
    
    \bottomrule
\end{tabular}
\tablefoot{$\mathcal{U}(a,\:b)$ indicates a uniform distribution between $a$ and $b$, $\mathcal{L}(a,\:b)$ a log-normal distribution, $\mathcal{N}(a,\:b)$ a normal distribution, and $\mathcal{T}(a,\:b)$ a truncated normal distribution (between 0 and 1) where $a$ and $b$ are the mean and standard deviation. The truncated Gaussian priors on TESS limb-darkening coefficients are converted from \citet{Claret2017}. $^{(*)}$~In the case of non-null eccentricity, the priors are set as follows: $(\sqrt{e}\,\sin\omega, \sqrt{e}\,\cos\omega)$ in $\mathcal{U}(-1.0,\:1.0)$.}
\end{table*}

\section{Additional plots}
\label{appendix:addplots}

\begin{figure*}
   \centering
   \includegraphics[width=0.8\textwidth]{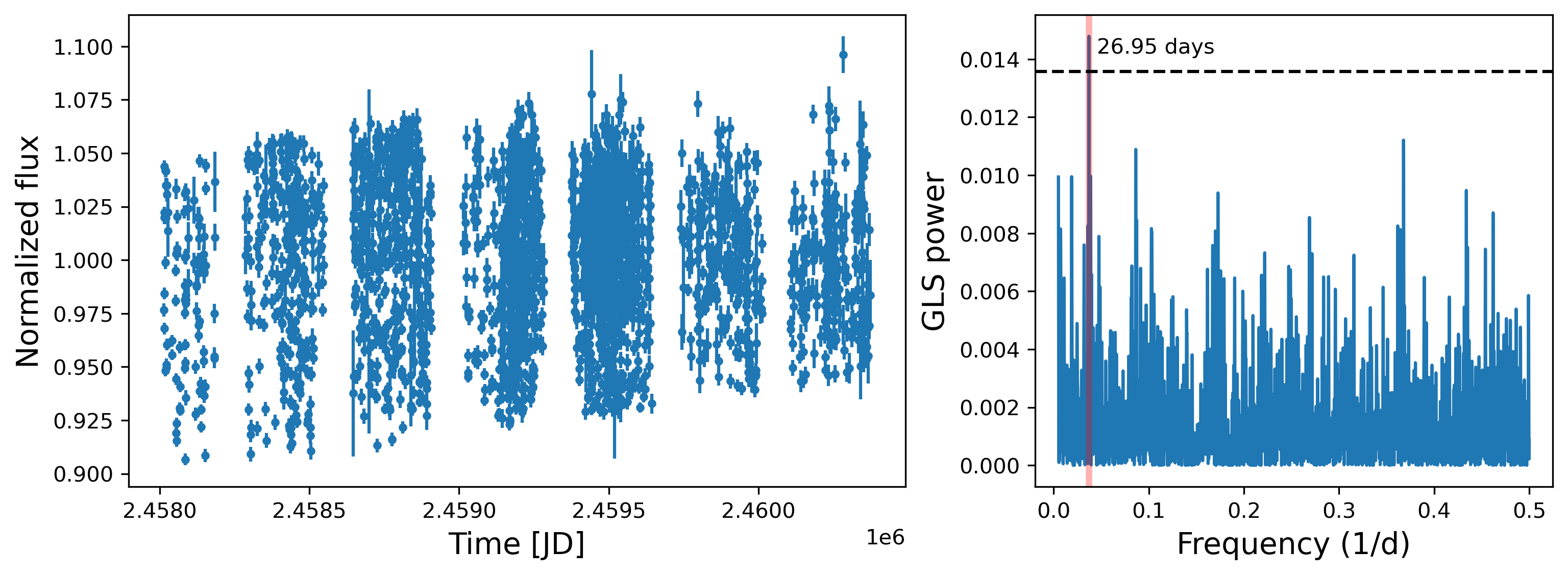}
      \caption{\textbf{Left panel}: The ASAS-SN timeseries for BD+00\,444 in the Sloan $g$-band, corrected for a linear trend. \textbf{Right panel}: The GLS periodogram of the time series, with its highest peak highlighted in red. The horizontal dashed line corresponds to a FAP of 0.1\% (evaluated with the bootstrap method).}
         \label{fig:asas}
\end{figure*}

\begin{figure*}
    \centering
    \includegraphics[width=\textwidth]{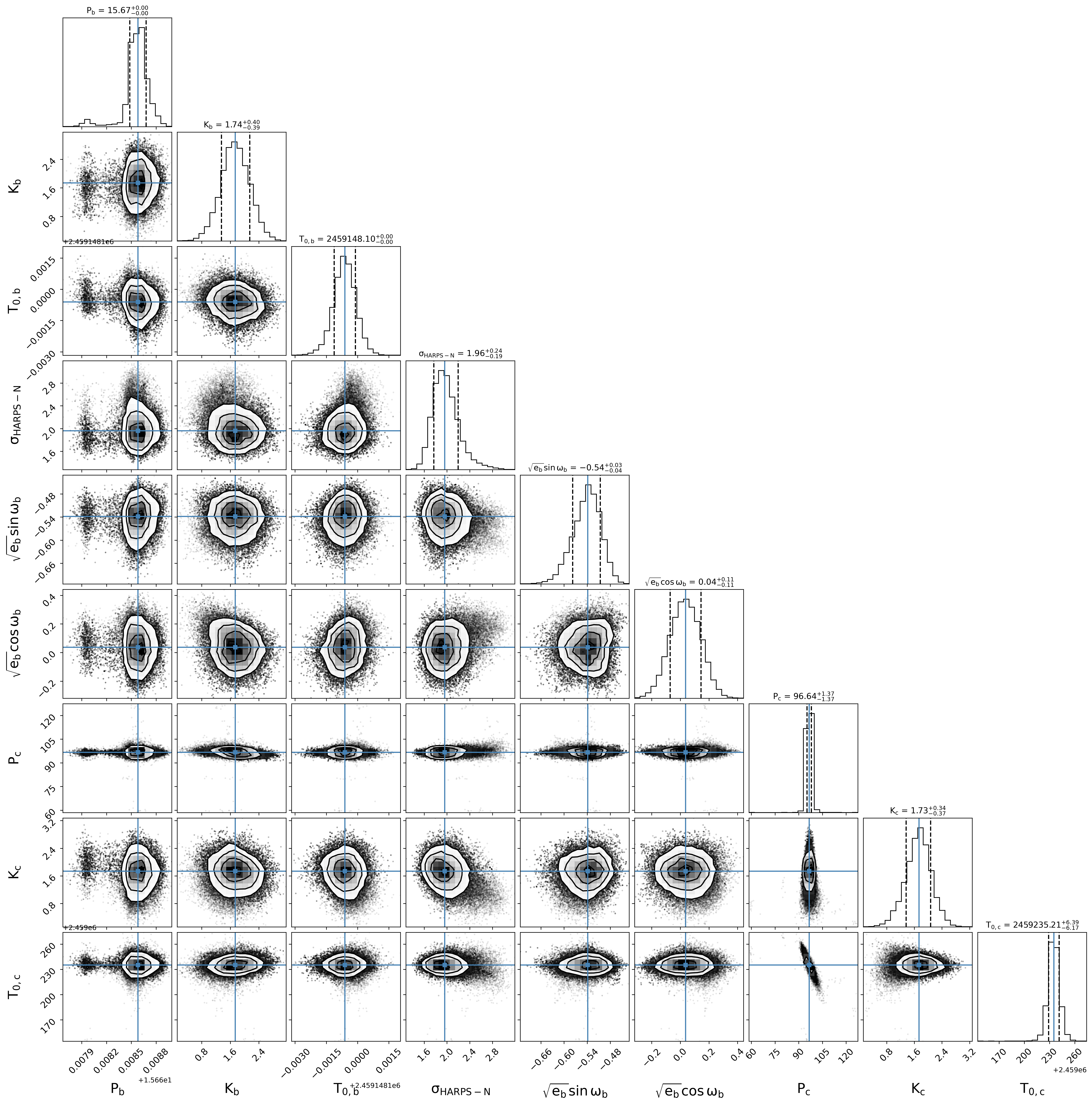}
    \caption{Corner plot of the posterior distribution for the preferred joint transit and RV analysis of Sect.~\ref{sec:joint_analysis}.}
    \label{fig:corner2p}
\end{figure*}

\end{document}